\documentclass{article}

\usepackage{arxiv}

\usepackage[utf8]{inputenc} 
\usepackage[T1]{fontenc}    
\usepackage{hyperref}       
\usepackage{url}            
\usepackage{booktabs}       
\usepackage{amsfonts}       
\usepackage{nicefrac}       
\usepackage{microtype}      
\usepackage{lipsum}

\usepackage{listings}
\usepackage{caption}

\DeclareCaptionFormat{mylst}{\hrule#1#2#3}
\usepackage{xcolor}
\usepackage{graphicx}
\usepackage{caption}
\usepackage{cite}
\usepackage{mathptmx}
\usepackage{amsmath}
\usepackage{amssymb}
\usepackage{amsthm}
\usepackage{rotating} 
\usepackage{array}
\usepackage{enumitem}
\usepackage[caption=false]{subfig}
\usepackage{multirow}

\usepackage{rotating}
\usepackage[title]{appendix}

\newcommand{\AdminNet}{Identity Blockchain Network}
\newcommand{\CTINet}{Activity Blockchain Network}
\newcommand{\methodName}{TRADE}
\newcommand{\tradeHeader}{TRADE Header}

\newcommand{\tuid}{pseudo-identity}
\newcommand{\authorizationPolicy}{authorization policy}
\newcommand{\badgePolicy}{badge policy}

\title{TRusted Anonymous Data Exchange: Threat Sharing Using Blockchain Technology}

\author{
  Yair Allouche \\
   IBM Cyber Security Center of Excellence \\
   Beer-Sheva 8410501, Israel \\
   \texttt{yair@il.ibm.com} \\

  \And
  
  Nachiket Tapas \\
    Department of Engineering \\
    Universita’ degli Studi di Messina \\
    Messina 98166, Italy \\
    \texttt{ntapas@unime.it} \\

  \And
  
  Francesco Longo \\
    Department of Engineering \\
    Universita’ degli Studi di Messina \\
    Messina 98166, Italy \\
    \texttt{flongo@unime.it} \\
  \And
  
  Asaf Shabtai \\
   Department of Software and Information \\
    Systems Engineering \\
   Ben-Gurion University of the Negev \\
   Beer-Sheva 8410501, Israel \\
   \texttt{shabtaia@bgu.ac.il} \\
  \And
  
  Yaron Wolfsthal \\
   IBM Cyber Security Center of Excellence \\
   Beer-Sheva 8410501, Israel \\
   \texttt{wolfstal@il.ibm.com} \\
}

\begin{document}
\maketitle

\begin{abstract}
Cyber attacks are becoming more frequent and sophisticated, introducing significant challenges for organizations to protect their systems and data from threat actors. 
Today, threat actors are highly motivated, persistent, and well-founded and operate in a coordinated manner to commit a diversity of attacks using various sophisticated tactics, techniques, and procedures. 
Given the risks these threats present, it has become clear that organizations need to collaborate and share cyber threat information (CTI) and use it to improve their security posture. 
In this paper, we present {\methodName} -- TRusted Anonymous Data Exchange -- a collaborative, distributed, trusted, and anonymized CTI sharing platform based on blockchain technology. 
{\methodName} uses a blockchain-based access control framework designed to provide essential features and requirements to incentivize and encourage organizations to share threat intelligence information. 
In {\methodName}, organizations can fully control their data by defining sharing policies enforced by smart contracts used to control and manage CTI sharing in the network. 
{\methodName} allows organizations to preserve their anonymity while keeping organizations fully accountable for their action in the network. 
Finally, {\methodName} can be easily integrated within existing threat intelligence exchange protocols - such as trusted automated exchange of intelligence information (TAXII) and OpenDXL, thereby allowing a fast and smooth technology adaptation.
\end{abstract}


\section{\label{sec:intro}Introduction}
Cyber-security tops the priority list among businesses, irrespective of their size~\cite{travelers2019}.
Nearly 80\% of respondents attributed it to rapid advances in technology-driven by the Internet of Things (IoT) devices, mobility, cloud-based workloads, and SaaS, leading to an increase in the cyberattack surface.
Cyber attacks are becoming more frequent and sophisticated, introducing ever-increasing challenges to organizations, protecting their systems and data from threat actors~\cite{platt2018}. 
These days, threat actors are often highly motivated, persistent, and well-funded, operating in coordinated groups either as part of a criminal enterprise or on behalf of a nation-state~\cite{accenture2019}. 
Threat actors can use various tactics, techniques, and procedures (TTP) to commit attacks, resulting in financial fraud, disruption of services, system compromise, or theft of sensitive information~\cite{lemay2018survey}.

To deal with the risks presented by cyber threats, organizations need to work together in a coordinated fashion to share cyberthreat information (CTI) and thereby improving their security posture~\cite{brown2019evolution}.
By exchanging CTI, organizations gain a more comprehensive understanding of the threats they may face by taking advantage of the sharing community's collective knowledge, capabilities, and experience. 
For example, an organization might improve its security posture by using shared information to build threat-informed detection, defense, and mitigation strategies. 
By correlating CTI from multiple sources, an organization enriches and validates existing information. 
The correlation, in turn, makes the information more reliable and actionable and reduces the exposure time and false alarms~\cite{settanni2017acquiring}. 
The sharing of CTI allows organizations to detect better cyber attack campaigns that target particular industry sectors, business entities, or institutions~\cite{tounsi2018survey}.
Additionally, organizations that receive CTI and subsequently use this information to remediate a threat can offer a degree of protection to other similar organizations by impeding the threat's ability to spread.
Though the benefits of threat sharing are well known, the practice of cyber threat sharing is still limited. 

According to NIST special publication guide to CTI sharing~\cite{johnson2016guide}, the critical barriers to extensive and effective threat sharing include: establishing trust relationships, achieving interoperability and automation, safeguarding sensitive information, protecting classified information, enabling information consumption and publication, accessing external information, evaluating the quality of received information, complying with legal and organizational requirements, and limiting attribution.
Trust formation is an essential characteristic of any information-sharing system.
Blockchain supporting distributed architecture, transparency, and accountability can ensure trust in the system.
Publicly available immutable logs with consensus engine enable blockchain to act as a policy and access control engine for an information-sharing network.
However, since the blockchain supports transparency, it does not support privacy and anonymity out of the box.
Also, guaranteeing privacy and anonymity limits the transparency and accountability of a blockchain-based system.
A novel mechanism is required to establish a trade-off between privacy, anonymity, transparency, and accountability.

In this paper, we present {\methodName} --- \textbf{TR}usted \textbf{A}nonymous \textbf{D}ata \textbf{E}xchange --- a blockchain-based threat intelligence sharing solution that allows peer-to-peer threat sharing. 
{\methodName} encourages sharing threat information by establishing producer/ consumer trust on the threat sharing network.
A trusted third party~\cite{fsisac} models and personal trust relationship-based peer-to-peer models ensure trust in threat intelligence sharing communities. 
However, trust in an external distributor and bottleneck introduced by the trusted central party is the model's challenges.
Also, the peer-to-peer model suffers from coverage and scalability issues.
Using blockchain permits {\methodName} to mimic the peer-to-peer trust model without coverage gaps and delays.
Each time a member of the network contributes, accesses, or enriches threat information, the transaction is recorded on the blockchain. 
This way, a full history of the information flow is immutably recorded and can be audited if necessary later.
A significant challenge related to Blockchain is the cost involved in the implementation.
However, considering the highly regulated industries like banks, the cost is deemed secondary compared to the Blockchain's security guarantees.

{\methodName} allows organizations to control who has access to their CTI (without revealing the source of information) and assess the quality of the threat information they consume.
Smart contracts are leveraged to define the levels of trust and anonymity required to enable collaboration. 
They enforce organizational requirements for attributes such as reputation and contribution levels to limit who has access to the threat intelligence information that members publish.

At the same time, the platform ensures transparency by ensuring that the network participants are held accountable for their actions.
{\methodName} introduces the role of Registrars, a trusted third party responsible for anonymizing the identity of the organizations on the network, among other actions.
The registrars form a multi-trusted third party that needs consensus to take action on the sharing network.
Thus, the sharing network ensures anonymity; at the same time, the registrars ensure accountability.

{\methodName} enables organic network formation in a bottom-up manner.
An organization wanting to share threat information can set up the platform and start sharing intelligence.
Like-minded organizations can form a coalition and share threat information.
Finally, {\methodName} leverages existing threat intelligence exchange protocols - such as trusted automated exchange of intelligence information (TAXII) and structured threat information expression (STIX) - which are integrated into industrial security workflows.
Thus, organizations with existing threat-sharing infrastructure can easily integrate {\methodName} into their threat-sharing workflow with minimum effort.

\noindent In summary, this paper provides the following contributions: (i) a blockchain-based CTI sharing framework that can be easily integrated with existing threat sharing infrastructures, (ii) a novel privacy-preserving network architecture that ensures anonymity, transparency, and accountability, (iii) a detailed description of smart contracts which support the operation of the network, ensuring security guarantees for CTI sharing, and (iv) a high-level description of the additional features to encourage the user to share.
\section{\label{sec:related}Related Work}
Cyber-attacks are quickly evolving into advanced persistent threats (APTs)~\cite{farwell2011stuxnet} via cooperation, isolation, and sophistication.
Consequently, new, more robust, and collaborative paradigms are required to defeat these kinds of attacks~\cite{virvilis2013big}, leading to situational awareness~\cite{jajodia2009cyber}. 
This section details the related work in the area of threat intelligence sharing and access control using blockchain, which enables CTI sharing. 
A summary of the related works and their evaluation against the requirements listed in \cite{johnson2016guide} is presented in Table \ref{tab:sumrelated}.

\begin{table}
\centering
\small
\caption{Summary of related works.}
\label{tab:sumrelated}
\begin{tabular}{|p{1.5cm}|p{3.3cm}|c|c|c|c|c|c|c|} 

\hline
Reference & Description & Distributed/ & Anonymity & Transpa- & Account- & Integration & Data \\
& & Centralized & & rency & ability & & Control \\
\hline
\cite{uscert} & Alerts generated by computer emergency response team & Centralized & -- & X & -- & -- & -- \\
\hline
\cite{mel2007nvd} & National vulnerability database alerts & Centralized & -- & X & -- & -- & -- \\
\hline
\cite{mell2007nist} & Common Vulnerability Scoring System (CVSS) & Centralized & -- & X & -- & -- & -- \\
\hline
\cite{cwemitre} & Common Weakness Enumeration (CWE) & Centralized & -- & X & -- & -- & -- \\
\hline
\cite{cve} & Common Vulnerabilities and Exposures & Centralized & -- & X & -- & -- & -- \\
\hline
\cite{btsecurities} & BT security threat monitoring & Centralized & X & X & -- & -- & X \\
\hline
\cite{cif} & Coco-Cloud CTI management system & Centralized & X & X & -- & -- & X \\
\hline
\cite{shu2015privacy} & Hybbrid CIDS based framework & Centralized & X & -- & -- & -- & X \\
\hline
\cite{fung2010privacy} & Privacy preserving CTI sharing framework & Centralized & X & -- & -- & -- & -- \\
\hline
\cite{chadwick2020cloud} & Cloud based CTI sharing framework & Centralized & X & -- & -- & X & -- \\
\hline
\cite{protective} & Proactive Risk Management through Improved Cyber Situational Awareness & Decentralized & X & X & -- & -- & -- \\
\hline
\cite{tradelens} & TradeLens: blockchain based supply chain tracking & Decentralized & X & X & -- & -- & -- \\
\hline
\hline
Proposed Solution & TRADE: Blockchain based CTI sharing platform & Decentralized & X & X & X & X & X \\
\hline
\end{tabular}
\end{table}

\subsection{CTI sharing platforms}
\label{ctirelated}
Some of the initial works addressing threat sharing techniques, including indicators of compromise and standard sharing format, are presented in~\cite{moriarty2011incident}.
Publishing security alerts publicly \cite{uscert}, national vulnerability database alerts \cite{mel2007nvd}, and security announcements by security vendors are some of the traditional methods of sharing cyber threat intelligence.    
In recent years, more advanced methods like common vulnerability scoring system \cite{mell2007nist}, common weakness enumeration \cite{cwemitre}, and common vulnerabilities and exposures \cite{cve} notifications are adopted. 
However, all the approaches mentioned above are unidirectional and do not encourage cooperation among organizations. 
BT security threat monitoring \cite{btsecurities}, Coco-Cloud \cite{caimi2015legal}, and CIF \cite{cif} aggregate and share threat information privately.
The proposed solutions do not support encryption or anonymization to protect sensitive CTI information.

Public domain projects (PROTECTIVE \cite{protective}) and industry-focused collaborations (FS-ISAC \cite{fsisac}, and R-CISC \cite{rcisc})
have gained popularity in recent years.
However, CTI sharing is happening either manually or in a supervised fashion.
A cloud-native and blockchain-based framework is presented in Tradelens \cite{tradelens}.
Though used for logistics use-case, the underlying framework can be used for any information exchange platform.
The framework enables anonymity and trust by allowing information exchange between known entities.
However, the information is present on a single node and is shared with other entities making it vulnerable to a single point of failure.

\subsection{Access control using blockchain}
\label{accessrelated}
Blockchain-based access control tries to address a trusted and transparent third-party requirement by leveraging smart contracts.
FairAccess~\cite{ouaddah2016fairaccess} use blockchain transactions to allow, delegate, and revoke access to the resources.
However, mining latency and the size of the underlying blockchain platform limit the wide adoption.
A similar proposal is presented in \cite{novo2018blockchain}.
The proposed architecture stores the access control policy for the IoT devices on the blockchain and can be accessed globally for verification.
\cite{tapas2018blockchain} addresses the problem of mining latency by taking advantage of the transaction pool.
A similar approach is presented in~\cite{cruz2018rbac, ferretti2019authorization}, which ensures users with the role can access the resources in a challenge-response authentication protocol.
An XACML-based access control system is presented in~\cite{maesa2019blockchain}. 
The proposal uses Ethereum blockchain smart contracts to evaluate the access request against the security policy.
In contrast, authors in~\cite{kokoris2018hidden} present a platform using blockchain as an immutable log of operations.
The immutable log ensures the participants audit the operations and improves the trust in the system.
Similarly, in~\cite{ali2019blockchain}, the authors propose blockchain as the enabler of access control for IoT.
An attribute-based access control scheme is proposed in~\cite{zhang2020attribute}, which uses blockchain for authorizing IoT devices.
Trusted entities, called authority nodes, are used to interact with the blockchain and authorize access.
TrustAccess~\cite{gao2020trustaccess} presents another attribute-based access control that addresses the privacy of attributes and policies using ElGamal homomorphic encryption (to ensure the privacy of attribute during authorization validation).

The blockchain-based solutions present in the literature suffer from the problem of privacy.
Since the information stored on the blockchain is publicly available to the participants to ensure auditability and transparency, this leads to loss of privacy.
The solutions addressing the privacy issue usually depend on encryption to protect the data from unauthorized access.
The dependency makes the solution rigid and static, as information like the participants interested in accessing the information needs to be available in advance.
Our proposed solution separates identity and information to tackle these issues.
\section{\label{sec:proposed}Proposed Platform}
\begin{figure}[t]
\centering
\includegraphics[width=0.9\textwidth]{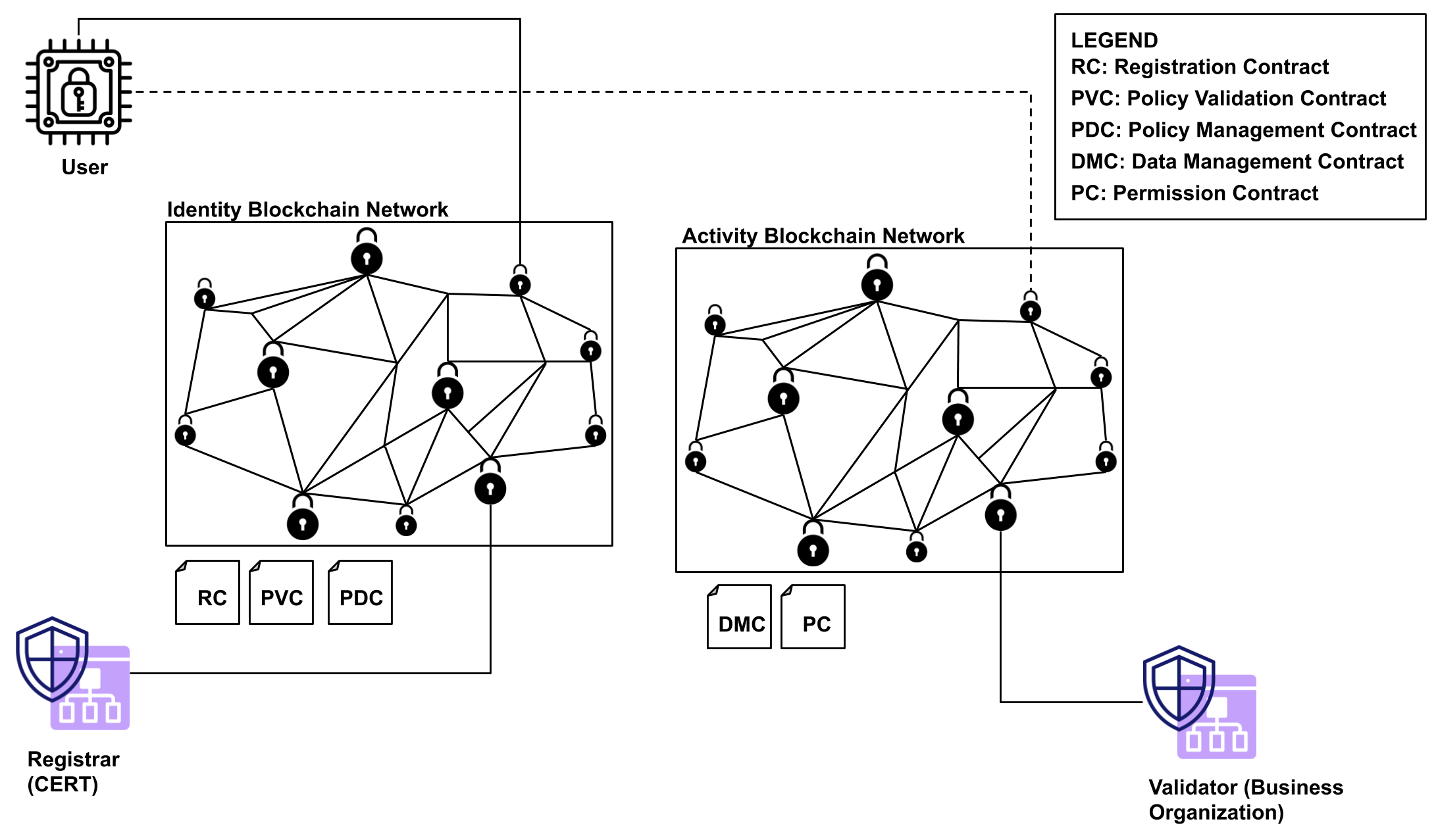}
\caption{\methodName~network architecture.}
\label{fig:netarch}
\end{figure}

We propose a permissioned blockchain-based threat sharing platform (see Figure~\ref{fig:netarch}) for representing and sharing access rights in a transparent and auditable manner.
The proposed platform stores the access policies on the blockchain and validates the access to a resource on the blockchain.
In this section, we introduce the basic building blocks of the proposed solution.

\subsection{\label{subsec:opennetwork}Open network}
To establish an effective cyber resilience against cyber threats, more organizations need to join and share cyber threat intelligence.
An \textit{open network} is an essential component to facilitate such sharing.
In an open network, any organization interested in sharing/consuming threat intelligence can interact with any other organization on the network.
Such an open network requires trust and access control, which needs to be achieved without depending on a single trusted third party. 

In this paper, we suggest applying a blockchain-based access control~\cite{rouhani2019blockchain} to address those challenges. 
Blockchain-based access control relieves the storage servers from the burden of handling the complex access control management in such complex business networks, and at the same time, mitigates the need for outsourcing these functionalities to a trusted, powerful entity. 
The business rules are enforced with smart contracts, which are stored on the blockchain and, thus, are auditable.
Also, nodes only need to verify the access token's validity, which is securely stored on the blockchain. 
The on-chain storage allows shifting those services to the cloud. 
Finally, blockchain-based access control provides user-driven, transparent access control. The data owner is the only one responsible for defining the granular access control policies over his data.

\subsection{\label{subsec:decoupled}Decoupled network architecture}
\label{subsec:dcplarch}
The threat sharing network consists of two permissioned blockchain networks: {\AdminNet} and {\CTINet} (see Figure~\ref{fig:netarch}).

The \textit{\AdminNet} consists of profiles and policies of the organizations interested in sharing/consuming the threat information.
Since the information present on this network is confidential and critical, access to the {\AdminNet} is regulated by trusted entities called \textit{Registrars}.
Organizations like US CERT \cite{Spafford88, Scherlis88}, US ISAC \cite{clinton1998presidential} or other well-reputed organization can act as registrars on the {\AdminNet}.
The registrars on the {\AdminNet} may form a consortium to ensure transparency on the network and regulate access to the threat sharing network. 
The opportunity to run independent validator nodes increases the security and reliability of the TRADE consensus protocol by having non-correlated failure risks.
The policies to share and access the threat information and smart contracts to validate the access request are stored on the {\AdminNet}.
The smart contracts handling the profile and policy management are stored on {\AdminNet}.

The \textit{\CTINet} logs and monitors the threat-sharing activities related to an organization.
The network holds two kinds of information.
The metadata related to the threat information is stored on the blockchain in the form of transactions.
The metadata acts as a history of records for auditing and helps in searching and generating notifications related to the availability of the threat information. 
Also, the information related to access-policies is stored on the blockchain.
An organization's access request is validated using the {\AdminNet}, and access permission is stored on the blockchain on successful validation.
The on-chain storage is done to eliminate any modification to the access token and increase trust in the network.
The smart contracts handling the data management and policy validation are stored on {\CTINet}.

Table \ref{tab:allowtran} summarizes the access authorizations of various entities of the system.
The decoupling of the {\AdminNet} from the {\CTINet} preserves privacy and anonymity while ensuring accountability.
After registering the organization on {\AdminNet}, a pseudo-identity is created for the organization.
The pseudo-identity protects the organization's actual identity and anonymizes the organization's operations on the {\CTINet}.
The mapping between the real identity and the pseudo-identity is stored off-chain to hold the organizations accountable for their actions on the {\CTINet}.

\subsection{Policy-based access control for threat sharing}
\label{subsec:policy}
Traditional access control schemes (such as attribute-based access control and role-based access control) are not suitable for threat-sharing scenarios. 
They have a built-in bias towards a static approach that is increasingly ineffective within the dynamic systems within which access decisions need to be made.
The proposed solution uses policy-based access control (PBAC), which uses context-based security policies. 
Essentially, PBAC combines attributes from the resource, the environment, and the requester with information on the particular set of circumstances under which the access request is made and uses rule sets that specify whether the access is allowed under the organizational policy attributes under those circumstances. 
This means that a policy-driven workflow engine discovers, organizes, and resolves information or attributes to provide the necessary context for more informed decision-making about access. 

In PBAC, each organization could use different sharing policies that might reflect different trust levels. 
When sharing CTI, the organization might tag the shared CTI with the policy to be enforced when disseminating the CTI within the network. 
For example, an organization might apply a range of sharing policies with varying degrees of restriction. 
When sharing highly confidential CTI, the organization would tag the CTI with a highly restrictive policy. 
Only the network members that satisfy this policy would be able to consume this sensitive data. 
However, when an organization wishes to share less sensitive CTI, it might tag it with a less strict sharing policy. 
Any organization meeting those less strict policy guidelines could then consume it.

\subsection{Smart contracts}
\label{subsec:smartcontracts}
\begin{figure}[t]
\centering
\includegraphics[width=1\textwidth]{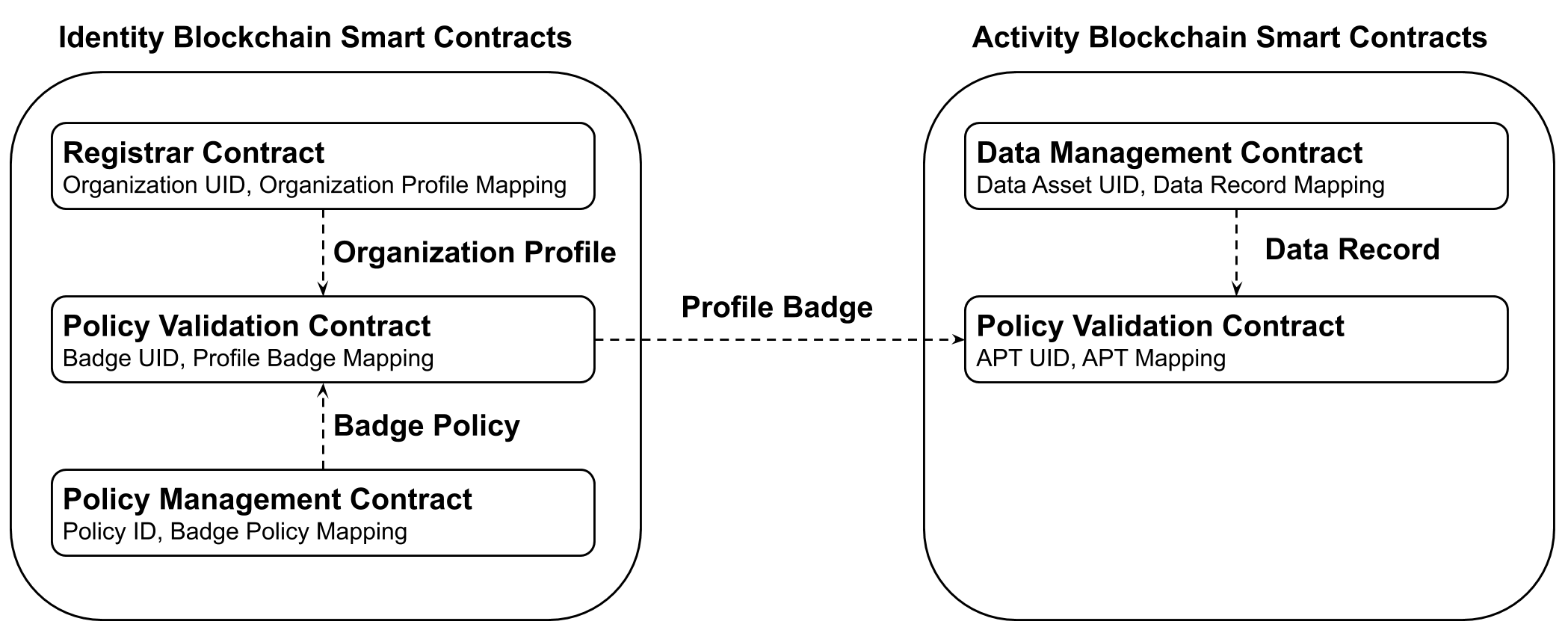}
\caption{Smart contracts on the two blockchain networks.}
\label{fig:smartcontracts}
\end{figure}

The access and security policies in the framework are enforced using smart contracts. 
The following contracts are designed to govern CTI exchange among organizations within a threat information-sharing network. 
To accomplish this, the contracts contain metadata about record ownership, access permissions, and data integrity. 
The blockchain transactions in such a system carry cryptographically signed instructions to manage these properties. 
The contract’s state-transition functions carry out policies, enforcing data altercation only by legitimate transactions. 
Such a policy could be designed to implement any set of rules governing a particular CTI record, given it could be computationally defined. 
The proposed system locates these policies on a blockchain through the following five smart contracts (see Figure \ref{fig:smartcontracts}).

\noindent \textbf{Registration contract (RC).}
The registration contract regulates the organization profile management on the \AdminNet. 
It maps an organizations' \textit{permanent} blockchain address identity (equivalent to a public key) to the organization profile. 
An organization profile is a set of organizational attributes used by a policy validation contract (PVC) to validate whether an organization fulfills network members' policies. 
The organization profile may include various attributes such as an organization's size, headquarter location, credit rate, and whether the organization is GDPR compliant. 
An RC contract would regulate the registration of new identities. 
Identity registration would be restricted only to certified institutions referred to as registrar, which regulates adding a new organization to the network, i.e., adding a mapping between organizations' blockchain address identity to the organization profile. 
Note that by adding this mapping, the registrar is not vouching for the organization but ensures that the organization profile genuinely reflects the organizational attributes. 
When adding a new organization, the registrar stores off-chain a mapping between the real organization identity and its blockchain address identity. 
A set of public keys is stored as part of the organizations' profile and is used to authenticate the organization. 
By employing separate key sets and separate transactions, this approach allows organizations to maintain a high degree of anonymity. 
The RC contract is also responsible for addressing network policy violations and resolve a dispute among network members. 
In case of violation, the RC defines the terms at which the registrar exposes the organization's real identity. 
For example, when more than x registrars have determined that this organization has violated the network policy. 
The RC contract also defines the terms at which organizations are revoked, or a violation note is added to the organization profile.

\begin{table}
\centering
\small
\caption{Entities and their allowed operations and transactions on \AdminNet~and \CTINet.}
\label{tab:allowtran}
\begin{tabular}{|p{1.5cm}|p{2cm}|p{2cm}|p{2cm}|p{2cm}|}
\hline
 \multirow{2}{1.5cm}{\textbf{Entities}} & \multicolumn{2}{c|}{\textbf{Identity Blockchain Network}} & \multicolumn{2}{c|}{\textbf{Activity Blockchain Network}}\\
 \cline{2-5}
 & \multicolumn{1}{|c|}{\textbf{Operations}} & \multicolumn{1}{|c|}{\textbf{Allowed Transactions}} & \multicolumn{1}{|c|}{\textbf{Operations}} & \multicolumn{1}{|c|}{\textbf{Allowed Transactions}} \\
\hline
Registrar & \multicolumn{1}{|c|}{NA} & 
\multicolumn{1}{|l|}{1. Identity Registration}
& \multicolumn{1}{|c|}{NA} & \multicolumn{1}{|c|}{NA} \\ 
& & \multicolumn{1}{|l|}{Transaction} & & \\ 

& & \multicolumn{1}{|l|}{2. Organization Revocation} & & \\ 
& & \multicolumn{1}{|l|}{Transaction} & & \\    

& & \multicolumn{1}{|l|}{3. Violation Addition}
& & \\
& & \multicolumn{1}{|l|}{Transaction}
& & \\

\hline

TRADE & \multicolumn{1}{|l|}{1. Access Permission} 
& \multicolumn{1}{|l|}{1. Profile Badge Request}
& \multicolumn{1}{|l|}{1. Notification}
& \multicolumn{1}{|l|}{1. Access Permission} \\

Client & \multicolumn{1}{|l|}{Request} 
& \multicolumn{1}{|l|}{Transaction}
& \multicolumn{1}{|l|}{2. Navigation}
& \multicolumn{1}{|l|}{Request (APR)} \\

& \multicolumn{1}{|l|}{2. Data Authentication}
& \multicolumn{1}{|l|}{2. Policy Creation}
& \multicolumn{1}{|l|}{3. Search}
& \multicolumn{1}{|l|}{Transaction}\\

& & & \multicolumn{1}{|l|}{4. Transparency} & \\

\hline

TRADE & \multicolumn{1}{|c|}{NA} & \multicolumn{1}{|c|}{NA} &
\multicolumn{1}{|l|}{1. Access Authorization} &
\multicolumn{1}{|l|}{1. Publish Data} \\

Server &  &  & \multicolumn{1}{|l|}{2. Data Authentication} &
\multicolumn{1}{|l|}{Transaction} \\

 &  &  &  & \multicolumn{1}{|l|}{2. Access Data} \\

 &  &  &  & \multicolumn{1}{|l|}{Transaction} \\

\hline

\end{tabular}
\end{table}

\noindent \textbf{Policy management contract (PDC).}
The policy management contract regulates the deployment and management of \textit{Profile Badge} policies on the {\AdminNet}. 
A profile badge policy is a set of business rules that need to be fulfilled by the organization profile to grant this organization with a profile badge. 
As in~\cite{maesa2018blockchain}, we suggest expressing the profile badge policy through attribute-based access control (ABAC) policies. 
An attribute-based access control policy combines rules expressing conditions over a set of attributes paired to the subject, the resource, or the environment. 
One of the key advantages of the ABAC languages (such as XACML \cite{anderson2003extensible}) is the richness of the language, where rules are conjunctively or disjunctively combined. 
Each policy is associated with a policy owner where only the policy owner can delete or modify the policy. 
However, the policy owner can grant other organizations to use his deployed policies. 
In section \ref{sec:perf}, we describe how this can highly minimize the load on the {\AdminNet} and increase anonymity.

\noindent \textbf{Permission contract (PC).}
The permission contract regulates the access permission tokens (APTs) generation on the {\CTINet}. 
An APT is generated by the access permission request transaction, which takes as an input (a) the data asset UID, (b) the desired access privilege, and (c) a set of profile badge references. 
Upon an access permission request transaction, the PC retrieves the corresponding data record from the data management contract using the data asset UID and extracts the authorization policy for the desired privilege. 
Next, the contract retrieves all the profile badges, using the set of profile badge references specified in the request, from the PVC on the {\AdminNet}. 
The contract would generate an APT if the retrieved profile badges satisfied the authentication policy. 
The APT record includes the following attributes (a) the requester’s public key, (b) access permission privilege, and (c) expiration time. 
Once the APT is stored within the blockchain, a pointer to this APT is returned to the access permission request. 
When the organization is communicating with the CTI server, it provides the pointer to the APT. 
The CTI server queries the blockchain with the given APT pointer to retrieve the APT record. 
Using the APT, the CTI server can grant access to the organization following the privilege defined in the APT. 
The PC contract allows the data owner to query the APTs corresponding to his data to provide the data owner full transparency over his data.

\noindent \textbf{Policy validator contract (PVC).}
The policy validator contract regulates the profile badge management on the {\AdminNet}. 
It maps an organizations' temporary blockchain address identity to profile badge. 
The Permission Contract uses the profile badge for generating Access Permission Tokens on the {\CTINet} for the anonymous temporary blockchain address identity. 
When an organization generates a profile badge for a given policy, it invokes a profile badge request transaction specifying the target policy ID and the temporary blockchain address identity it would like to use. 
This transaction triggers the PVC to validate the policy specified by the given policy ID against the requester's organization profile. 
In case the organization profile satisfied the policy, a profile badge is stored on the {\AdminNet}. 
The profile badge includes the following attributes: (a) badge ID, (b) the temporary public key, (c) the policy UID, and (d) validity flag. 
The profile badge revocation transaction deactivates the profile badge. 
As will be described in section \ref{sec:perf}, the reuse of profile badges by the organizations is desired to optimize the overall performance of TRADE. 
Hence, the profile badge's soundness stays as long as the organization profile has not been modified. 
Any profile update triggers the revocation of all its Profile Badges. 

\noindent \textbf{Data management contract (DMC).}
The data management contract regulates the creation and management of data records on the {\CTINet}. 
This contract stores a data record for any data asset in TRADE. 
A data asset can a CTI record or data set on a CTI server or data channel in a pub-sub CTI network. 
The data record includes the following attributes (a) data owner \textit{temporary} blockchain identity, (b) data access policy for the subject data (privilege mapping). 
Besides, the data record may also include (a) a hash of the data record that is used for authenticity validation of the data by the data consumers), (b) the data record description and keywords that are used for data search and notifications, and (c) communication information (server address) for record retrieval.
The DMC also supports search and notification mechanisms. 
The search mechanism allows organizations to retrieve data records by querying the data asset metadata (keywords, topic description). 
The notification mechanism allows organizations to be notified of network activity based on their interest. 
\section{Implementing {\methodName} with TAXII-based CTI Sharing Platform}
\label{sec:proposal}
We provide a detailed description of {\methodName} in which the CTI data exchange subnetwork is based on TAXII implementation~\cite{connolly2014trusted}. 
A similar integration approach can be used with other technologies such as OpenDXL and MiSP.
We begin by describing the interface between {\methodName} and the TAXII implementation (Section~\ref{subsec:interface}), and then we provide a detailed description of the access authorization process (Section~\ref{subsec:authorization}). 

\subsection{\label{subsec:interface}{\methodName} interface}
In this setup, the TAXII server delegates the access control processes to {\methodName}. 
The delegation relieves the TAXII servers from the burden of the complex access control management in such a complex business network and allows it to serve any member in {\methodName}. 
To provide simple integration of the existing CTI sharing technology, such as TAXII and OpenDXL, we suggest an SDK-based model. 
The TAXII clients and servers are equipped with the corresponding {\methodName} client and server SDKs. 
Since the {\methodName} SDKs hold the cryptography secrets, we suggest deploying them within a secure hardware module that is already available as a service in the cloud environment. 

\noindent The {\methodName} client SDK will provide the following services: 
\begin{itemize}
    \item \textbf{Access permission request:} 
    On access permission request, the {\methodName} client SDK returns a {\tradeHeader}, containing a reference to an \textit{access permission token} (APT) stored on the {\CTINet}. 
    The {\tradeHeader} is then added to the TAXII client request.
    \item \textbf{Data authentication:} 
    When a CTI producer uploads a new CTI record to the TAXII server, it affixes the data with its hash. 
    The CTI record consumers then use this hash to ensure that the data in question has not been altered.
    \item \textbf{Notification:} 
    The {\CTINet} can also act as a network coordinator using notifications. Specifically, the {\CTINet} can notify relevant organizations about a newly created record being added or modified or about creating a new trust group.
    \item \textbf{Navigation:} 
    When a CTI producer uploads a new CTI record to the TAXII server, it may affix the record with additional information indicating where the relevant record can be accessed in the network.
    \item \textbf{Search:} 
    As the network coordinator, the {\CTINet} would provide a CTI search mechanism for the network's contained records. 
    As the data itself is not stored on the blockchain when a CTI producer uploads a new data set to the TAXII server, he/she may mark the data with different tags describing the CTI record (e.g., financial malware, suspicious IP) to be used by the search mechanism. 
    \item \textbf{Transparency:} 
    This service allows network members to retrieve information related to their data stored on the Activity Blockchain ledger. 
    For example, this service allows a CTI producer to retrieve the history related to the download of its records.  
\end{itemize}

\noindent The {\methodName} server SDK provides the following services:
\begin{itemize}
\item \textbf{Access authorization:} 
    On access authorization request, the {\methodName} server SDK takes as the input the {\tradeHeader} produced by the client SDK, retrieves the APT from the {\CTINet}, and returns the privilege to the entitled requester.
\item \textbf{Data authentication:}
    Using this interface, the TAXII server can validate the authenticity of a received CTI record using the corresponding data record stored on the {\CTINet}. 
\end{itemize}

\subsection{\label{subsec:orgregistration}Organization registration process}
In this section, we describe the process of registration for a new organization interested in sharing cyber threat information. 
The organization presents its profile to the consortium of trusted third parties called registrars. 
The profile consists of information like (a) organization name, (b) the number of employees, (c) annual revenue, and (d) headquarter location.
The registrar validates the profile off-chain, and on successful verification, she saves the profile on the Identity Blockchain Network. 
Also, she creates a pseudo-identity on the Identity Blockchain Network for the organization to interact with the Activity Blockchain Network. 
The pseudo-identity profile consists of information like (a) public key, (b) identity tags, (c) CTI contribution, (d) status, and (e) issuer.
Once done, the registrar saves the mapping between the original identity and the pseudo-identity on her local database.
The off-chain storage of mapping ensures privacy and accountability as explained in Section \ref{subsec:decoupled}. 
Finally, she returns the blockchain address of the pseudo-identity to the organization. 

\subsection{\label{subsec:policycre}Consumption and sharing policy creation process}
In this section, we detail the following process in the sequence of events. 
Once the organization identities are created, the next step is to create the organization's consumption and sharing policies.
The consumption policy defines the organization's parameters before consuming the cyber threat information from a source.
It consists of (a) policy id, (b) description, and (c) terms of the policy.
The terms of the policy are a set of conditions that need to be satisfied for the policy's enforcement.
Similarly, the sharing policy defines the set of parameters that the organization looks for before sharing the cyber threat to a consumer. 
Thus, to successfully consume cyber threat information, the organization's consumption policy must agree with the sharing policy of the source.
The consumption and sharing policies also protect the system against free-loaders (empty consumption policy) and malicious CTI injectors (empty sharing policy).

\subsection{\label{subsec:cyberinsert}Cyber threat information insertion process}
In this section, we describe the process of collecting cyber threat information for sharing with the community.
Once the organization creates the consumption and sharing policies, the next step is to add the threat information to the Taxii server.
Also, a suitable sharing policy should be associated with the threat information.
Once done, a notification is sent to the community members, informing them about the availability of new information.

\subsection{\label{subsec:authorization}Access authorization process}
This section provides a detailed description of the access authorization process of data retrieval from a TAXII server.
A very similar process can be applied to other data access authorization processes, including a topic subscription in a pub/sub model or joining a trusted group.

\begin{figure}[t]
\centering
\includegraphics[width=0.90\textwidth]{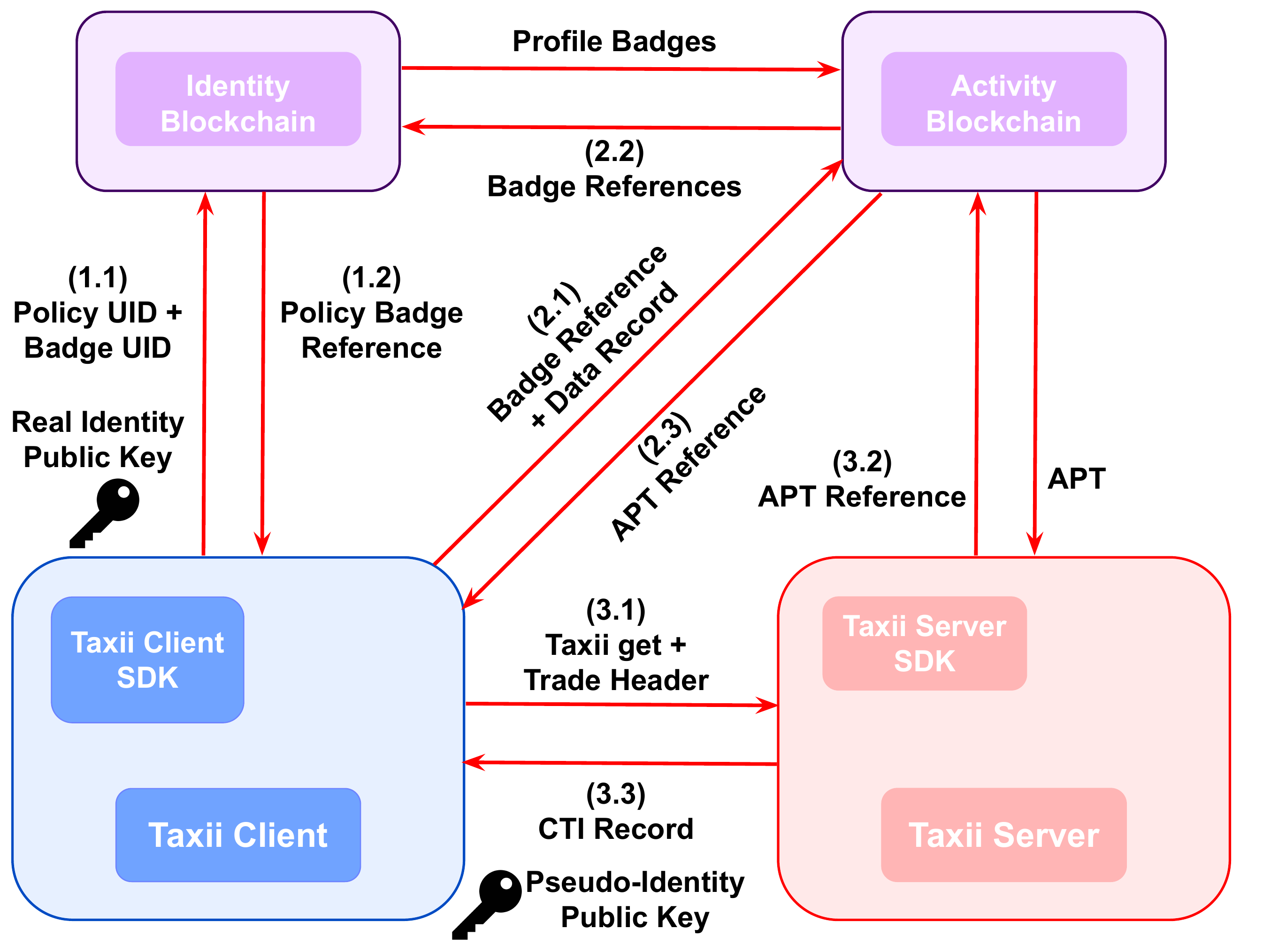}
\caption{Workflow of an authorization process.}
\label{fig:authorization}
\end{figure}

When the data producer $P$ uploads a new record to the TAXII server, he first invokes a \textit{publish data} transaction, which results in a new data record on the {\CTINet}. 
The data consumer can retrieve the data record by using the search and notification services of the {\methodName} client SDK (see Figure \ref{fig:authorization}). 
These services leverage the keywords provided by the data owner when data was uploaded. 
The data record provides the data consumer $C$ all the information he needs to retrieve the data record. 
Specifically, the data consumer $C$ extracts the following attribute from the data record: (a) record ID, (b) the {\authorizationPolicy}, (c) the record hash and optionally, and (d) the TAXII server address (access information). 

The data retrieval process includes the following three phases (see Figure~\ref{fig:authorization}):

\begin{description}
\item [Phase 1 -- Profile badge generation.]
In the first stage, the data consumer generates a Profile Badge record (if such does not exist) on the {\AdminNet} for each {\badgePolicy} specified in the record’s {\authorizationPolicy}. 
The Profile Badge includes the following attributes: (a) Badge ID, (b) the user {\tuid}, (c) the policy UID, and (d) Validity flag.
To generate the Profile Badges, the data consumer applies the following interaction with the {\AdminNet} using its real identity credentials. 

\begin{enumerate}[label=1.\arabic*]
    \item The data consumer first invokes a Profile Badge Request Transaction to the {\AdminNet} for each {\badgePolicy} specified in the record's {\authorizationPolicy}, which is not already stored on the blockchain.
    The transaction includes the policy ID and the data consumer's {\tuid}.
    \item The Policy Validator smart contract validates the policy specified by the given policy ID against the organization profile. 
    In case the organization profile satisfied the policy, the smart contract generates a Profile Badge (PB), which is stored on the {\AdminNet}. 
    In the end, a reference to the PB is returned to the requester.
\end{enumerate}

\item[Phase 2 -- Authorization token generation.]
In this stage, the data consumer generates an Authorization Token Record on the desired record's Activity Blockchain. 
To this end, the data consumer applies the following interaction with the Activity Blockchain using one of its temporary credentials.

\begin{enumerate}[label=2.\arabic*]
\item The data consumer invokes an Access Permission Request (APR) to the {\CTINet}, requesting access to a data item or an entire data set. 
The transaction includes the set of references of the relevant user badges (obtained at stage 1).
\item The RC smart contract retrieves the target PB from the {\AdminNet} using the PB reference provided by the user. 
In case all policies are valid (active), the RC smart contract generates an Access Permission Token (APT), which is stored on the blockchain. 
The stored APT would include the following: (a) the data consumer's public key, (b) access permission privilege, and (c) expiration time.
In the end, a reference to the APT is returned to the requester. 
To support flexible deployment, the response might also include additional information indicating the network location of the relevant database to be accessed (e.g., hostname and port in a standard network topology). 
\end{enumerate}

\item[Phase 3 -- Data retrieval.]
In this stage, the data consumer retrieves the TAXII server's data using the APT record on the {\CTINet}. 
To this end, the data consumer applies the following interaction with the TAXII server using one of its temporary credentials.

\begin{enumerate}[label=3.\arabic*]
\item The data consumer then sends a transaction request to the TAXII server, including (a) the requested operation and metadata required to apply the operation, (b) the user's signature created via the use of the user's private key, and (c) the reference to the APT provided by the {\CTINet}.

\item Upon receipt of such a transaction, the TAXII server would undertake the following actions:
(a) Given a valid timestamp, retrieve the target APT from the {\CTINet} using the data consumer's APT reference.
(b) Authenticate the user by using the user public key in the APT.
(c) Given a successful authentication of the requesting data consumer, validate the requested operation against the access permission privileges in the APT. 
(d) Given the requested operation conforms to the granted privileges in the APT, apply the operation, and return the relevant information to the requesting user.

\item After returning a record by a TAXII server to the requesting data consumer, the data consumer validates the data authenticity by using the corresponding hash from the data record. 
\end{enumerate}
\end{description}
\section{\label{sec:result}Formal Evaluation}

\begin{figure}[t]
\centering
\includegraphics[width=0.8\textwidth]{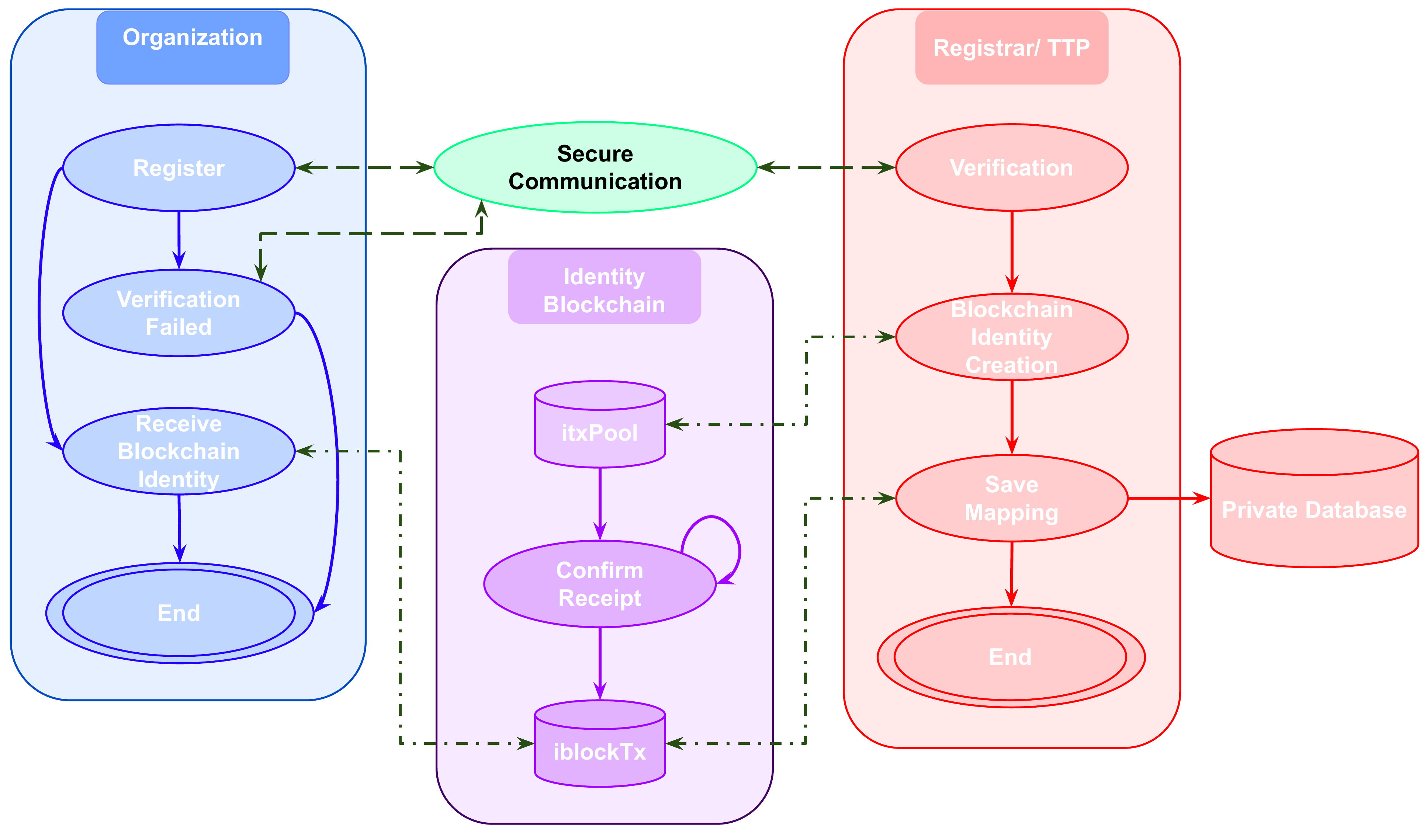}
\caption{Organization registration workflow.}
\label{fig:orgregistration}
\end{figure}

In this section, we formally evaluate the processes involved in the proposed framework. 
We begin by individually defining the previous section's processes as abstract specifications using TLA+ \cite{lamport2002specifying, lu2011towards} formal language for model verification.
Once the definitions are complete, the TLC model testing tool checks the proposal's correctness.
For the implementation of the framework, the resulting formal specification is then used as input.

\subsection{\label{subsec:orgspec}Organization registration specification}

In this section, we formalize the registration of a new organization in threat sharing community.
The workflow involves interaction between an organization (O) and a registrar (R) for adding the organization to the threat-sharing community (Figure \ref{fig:orgregistration}).
The two blockchains act as a ledger of records containing performed operations and are used as a communication medium.
We start with the initialization of each variable with the default value.
InitOrganization, InitRegistrar, and InitIdentityBlockchain are the organization, registrar, and identity blockchain variables' initial values.
The channel variable is used to describe safe communication between the organization and the registrar.
The state space for the registration workflow is defined as $Init \land \ast [\ Next ]\ _{vars} \land WF\_vars(Next)$ in TLA+ specifications where the $WF\_vars$ is for fairness.
The possible transitions from the initial state are described in the following equation:

\begin{minipage}{\linewidth}
\begin{lstlisting}[caption={State Space}]
 $Next\,\triangleq$ 
   $\lor\,\,Setup$
   $\lor\,\,\exists\,\,o \in Organization:$
     $\lor\,\,Preparation(o)\lor Register(o)$
     $\lor\,\,VerificationFailed(o)\lor ReceiveBlockchainIdentity(o)$
   $\lor\,\,\exists\,\,r \in Registrar$: 
     $\lor\,\,Verification(r)\lor BlockchainIdentityCreation(r)$
     $\lor\,\,SaveMapping(r)$
   $\lor\,\,ConfirmReceipt$
   $\lor\,\,Termination$
            
 $Spec\,\triangleq$
   $Init\,\land\,\ast\,[Next]_{vars}\,\land\,WF\_vars(Next)$
\end{lstlisting}
\end{minipage}

\textit{Next} defines a set of possible states for the system to transit. 
One of the states includes the \textit{Termination} state, which contains the system's conditions to terminate safely.
The channel variable is created to exchange messages between entities securely.

The organization is interested in sharing and consuming cyber threat information from the sharing network.
The organization begins with a $O\_Waiting$ and then transits through $O\_Register$, $O\_ReceiveBlockchainIdentity$, and $O\_Final$.
In the end, the organization receives a pseudo-identity reference which is stored on the identity blockchain.

\textit{(i) Register: }The Organization initiates the process by requesting the registrar to verify her profile and add her to the threat sharing network.
The profile information is sent via a channel to a registrar.
Once the information is sent, the system transits to \textit{O\_ReceiveBlockchainIdentity} state, where the organization waits for the blockchain address where her pseudo-identity is stored.

\begin{minipage}{\linewidth}
\begin{lstlisting}[caption={Register}]
   $\land\,\,oState[o]\,\,=\,\,O\_Register$
   $\land\,\,\exists\,\,r\,\,\in\,\,Registrar:$
     $\land\,\,rState[r]\,\,=\,\,R\_Waiting$
     $\land\,LET$
       $m\,\,==\,\,oBuffer[o]$
     $IN$
       $Send([src\,\,\mapsto\,\,o,\,\,dst\,\,\mapsto\,\,r,\,\,type\,\,\mapsto\,\,M\_Verification,\,\,data\,\,\mapsto\,\,\langle\,\,o,\,\,m[1],\,\,m[2]\,\,\rangle])$
   $\land\,\,oState'\,\,=\,\,[\,\,oState\,\,EXCEPT\,\,![o]\,\,=\,\,O\_ReceiveBlockchainIdentity\,\,]$
   $\land\,\,UNCHANGED\,\,\langle\,\,tempVars,\,\,oBuffer,\,\,registrarVars,\,\,identityBlockchainVars\,\,\rangle$
\end{lstlisting}
\end{minipage}

\textit{(ii) VerificationFailed: } If the organization's profile verification by the registrar fails, the system enters into \textit{O\_Finaly} state.
The system terminates safely with the organization not receiving the pseudo-identity details.

\begin{minipage}{\linewidth}
\begin{lstlisting}[caption={Verification Failed}]
   $\exists\,\,c\,\,\in\,\,channel:$
     $\land\,\,c.dst\,\,=\,\,o$
     $\land\,\,c.type\,\,=\,\,M\_NO$
     $\land\,\,oState'\,\,=\,\,[\,\,oState\,\,EXCEPT\,\,![o]\,\,=\,\,O\_Final\,\,]$
     $\land\,\,UNCHANGED\,\,\langle\,\,tempVars,\,\,oBuffer,\,\,registrarVars,\,\,commVars,\,\,identityBlockchainVars\,\,\rangle$
\end{lstlisting}
\end{minipage}

\textit{(iii) ReceiveBlockchainIdentity: }In this state, the organization waits for the message with pseudo-identity from the registrar.
Once the message is received, the organization is removed from the queue.
If more organizations need to register themselves, the process is repeated from the start with the system transiting to the \textit{O\_Register} state.
If no more registrations are required, the system moves safely to \textit{O\_Final} state and terminates.

\begin{minipage}{\linewidth}
\begin{lstlisting}[caption={Receive Blockchain Identity}]
   $\exists\,\,c\,\,\in\,\,channel:$
     $\land\,\,c.dst\,\,=\,\,o$
     $\land\,\,c.type\,\,=\,\,M\_Identity$
     $\land\,\,oState[o]\,\,=\,\,O\_ReceiveBlockchainIdentity$
     $\land\,\,IF\,\,Len(registrationPool)\,\,=\,\,0\,\,THEN$
       $\land\,\,oState'\,\,=\,\,[\,\,oState\,\,EXCEPT\,\,![o]\,\,=\,\,O\_Final\,\,]$
     $ELSE$
       $\land\,\,oState'\,\,=\,\,[\,\,oState\,\,EXCEPT\,\,![o]\,\,=\,\,O\_Register\,\,]$
   $\land\,\,UNCHANGED\,\,\langle\,\,tempVars,\,\,oBuffer,\,\,registrarVars,\,\,commVars,\,\,identityBlockchainVars\,\,\rangle$
\end{lstlisting}
\end{minipage}

The registrar is responsible for verifying the organization profile.
In case the verification is successful, the registrar creates a pseudo-identity on the identity blockchain and stores the mapping between the original identity and pseudo-identity.
The registrar begins with a $R\_Waiting$ and then transits through $R\_Verification$, $R\_BlockchainIdentityCreation$, and $R\_SaveMapping$.
Once the mapping is saved, and the pseudo-identity's blockchain address is sent back to the organization, the registrar moves back to $R\_Waiting$ state to process a new request.

\textit{(iv) Verification: }The registrar begins in \textit{R\_Waiting} state, waiting for the profile information from the organization.
Once the information is received, the registrar does an off-chain verification using the \textit{VerifyProfile} function, which abstracts the details.
If successful, the system transits to \textit{R\_BlockchainIdentityCreation} state.
Alternatively, the registrar sends a verification failure message back to the organization and transit to \textit{R\_Waiting}.

\begin{minipage}{\linewidth}
\begin{lstlisting}[caption={Verification}]
   $\exists\,\,c\,\,\in\,\,channel:$
     $\land\,\,c.dst\,\,=\,\,r$
     $\land\,\,c.type\,\,=\,\,M\_Verification$
     $\land\,\,rState[r]\,\,=\,\,R\_Waiting$
     $\land\,\,IF\,\,VerifyProfile(c.data[2],\,\,c.data[3])\,\,THEN$
       $\land\,\,rBuffer'\,\,=\,\,[\,\,rBuffer\,\,EXCEPT\,\,![r]\,\,=\,\,c.data\,\,]$
       $\land\,\,rState'\,\,=\,\,[\,\,rState\,\,EXCEPT\,\,![r]\,\,=\,\,R\_BlockchainIdentityCreation\,\,]$
       $\land\,\,UNCHANGED\,\,\langle\,\,tempVars,\,\,organizationVars,\,\,commVars,\,\,identityBlockchainVars\,\,\rangle$
     $ELSE$
       $\land\,\,rState'\,\,=\,\,[\,\,rState\,\,EXCEPT\,\,![r]\,\,=\,\,R\_Waiting\,\,]$
       $\land\,\,RecvInadditionSend(c,\,\,[src\,\,\mapsto\,\,r,\,\,dst\,\,\mapsto\,\,c.data[1],\,\,type\,\,\mapsto\,\,M\_NO,\,\,data\,\,\mapsto\,\,\langle\,\,\rangle])$
       $\land\,\,UNCHANGED\,\,\langle\,\,tempVars,\,\,organizationVars,\,\,rBuffer,\,\,identityBlockchainVars\,\,\rangle$
\end{lstlisting}
\end{minipage}

\textit{(v) BlockchainIdentityCreation: }With successful verification of the organization profile, the registrar creates a pseudo-identity of the organization on the identity blockchain.
The registrar sends a blockchain transaction with the profile details to the identity blockchain.
The organization can then use this identity to interact with the activity blockchain.
On successfully creating the identity on the blockchain, the system transits to \textit{R\_SaveMapping}.

\begin{minipage}{\linewidth}
\begin{lstlisting}[caption={Blockchain Identity Creation}]
   $\land\,\,rState[r]\,\,=\,\,R\_BlockchainIdentityCreation$
   $\land\,\,rState'\,\,=\,\,[\,\,rState\,\,EXCEPT\,\,![r]\,\,=\,\,R\_SaveMapping\,\,]$
   $\land\,\,blockchainIdentityTxn(r,\,\,BLOCKCHAIN\_ID,\,\,BLOCKCHAIN\_PROFILE)$
   $\land\,\,UNCHANGED\,\,\langle\,\,tempVars,\,\,organizationVars,\,\,rBuffer,\,\,commVars,\,\,iblockTx\,\,\rangle$
\end{lstlisting}
\end{minipage}

\textit{(vi) SaveMapping: }In this state, the registrar saves the mapping between the original identity and the pseudo-identity off-chain.
Once the mapping is saved, the registrar sends a successful message with the blockchain address back to the organization.
After the message is sent, the system transits back to \textit{R\_Waiting}.

\begin{minipage}{\linewidth}
\begin{lstlisting}[caption={Save Mapping}]
   $\land\,\,rState[r]\,\,=\,\,R\_SaveMapping$
   $\land\,\,LET$
     $data\,\,==\,\,mBuffer[r]$
   $IN$
     $Send([src\,\,\mapsto\,\,r,\,\,dst\,\,\mapsto\,\,data[1],\,\,type\,\,\mapsto\,\,M\_Identity,\,\,data\,\,\mapsto\,\,\langle\,\,BLOCKCHAIN\_ADDRESS\,\,\rangle])$
   $\land\,\,rState'\,\,=\,\,[\,\,rState\,\,EXCEPT\,\,![r]\,\,=\,\,R\_Waiting\,\,]$
   $\land\,\,UNCHANGED\,\,\langle\,\,tempVars,\,\,organizationVars,\,\,rBuffer,\,\,identityBlockchainVars\,\,\rangle$
\end{lstlisting}
\end{minipage}

\textit{(vii) ConfirmReceipt: }It is difficult to model an entire blockchain system. 
Thus, for our proposal, we model only its critical component. 
The blockchain framework is defined in our specification by independent ConfirmReceipt steps. 
The blockchain system confirms and transfers it to blockTx if there is a valid transaction, a compilation of verified transactions in the blocks.
In this case, \textit{TX\_ORIGINAL} and \textit{TX\_BLOCKCHAIN} are the valid transactions that indicate the original identity transaction and blockchain identity transaction.

\begin{minipage}{\linewidth}
\begin{lstlisting}[caption={Confirm Receipt}]
   $\exists\,\,receiptTx\,\,\in\,\,itxPool:$
     $\land\,\,(receiptTx.type\,\,=\,\,TX\_ORIGINAL\,\,\lor\,\,receiptTx.type\,\,=\,\,TX\_BLOCKCHAIN)$
     $\land\,\,itxPool'\,\,=\,\,itxPool\,\,\backslash\,\,\{receiptTx\}$
     $\land\,\,iblockTx'\,\,=\,\,iblockTx\,\,\cup\,\,\{receiptTx\}$
     $\land\,\,UNCHANGED\,\,\langle\,\,tempVars,\,\,organizationVars,\,\,registrarVars,\,\,commVars\,\,\rangle$
\end{lstlisting}
\end{minipage}

\textit{(viii) Termination: }If the organization receives the blockchain address of the pseudo-identity or the profile verification failure message, the entire process is terminated. 
Since no further changes are required, termination has no primed variables.

\begin{minipage}{\linewidth}
\begin{lstlisting}[caption={Termination}]
   $\land\,\,\forall\,\,o\,\,\in\,\,Organization\,\,:\,\,oState[o]\,\,=\,\,O\_Final$
   $\land\,\,UNCHANGED\,\,vars$
\end{lstlisting}
\end{minipage}

\begin{figure}[t]
\centering
\includegraphics[width=0.7\textwidth]{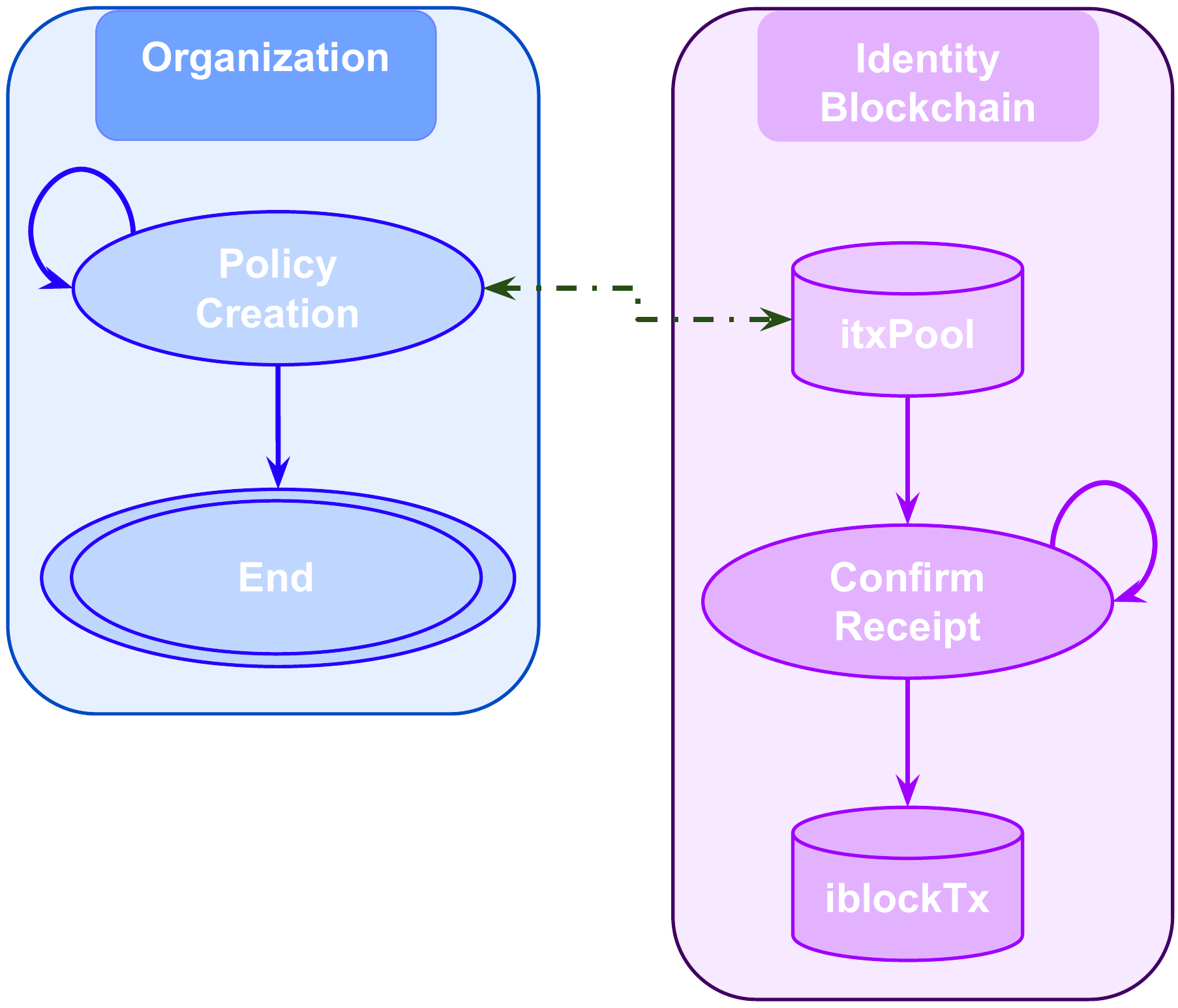}
\caption{Consumption and sharing policy creation Workflow.}
\label{fig:policycreation}
\end{figure}

\subsection{\label{subsec:policyspec}Consumption and sharing policy creation specification}Next, we formalize the creation of the consumption and sharing policies by the organization.
The workflow involves interaction between an organization (O) and the identity blockchain (Figure \ref{fig:policycreation}).
InitOrganization and InitIdentityBlockchain are the organization and identity blockchain variables' initial values.
The channel variable remains the same.
The state-space specification, too, for the policy creation workflow remains the same.
The possible transitions from the initial state are described in the following equation:

\begin{minipage}{\linewidth}
\begin{lstlisting}[caption={State Space}]
 $Next\,\triangleq$ 
   $\lor\,\,Setup$
   $\lor\,\,\exists\,\,o \in Organization:$
     $\lor\,\,Preparation(o)\lor CreatePolicy(o)$
   $\lor\,\,ConfirmReceipt$
   $\lor\,\,Termination$
            
 $Spec\,\triangleq$
   $Init\,\land\,\ast\,[Next]_{vars}\,\land\,WF\_vars(Next)$
\end{lstlisting}
\end{minipage}

In this case, the organization intends to define the consumption and sharing policies to be used for sharing and consuming cyber threat information. 
The organization begins with a \textit{O\_Waiting} and then transits through \textit{O\_CreatePolicy}, and \textit{O\_Final}. In the end, the organization stores all the policies on the identity blockchain.

\textit{(i) Preparation: }This state acts as a pre-processing state.
The system checks if there are any more policy creation requests in the queue.
If the queue is empty, the system transits to the \textit{O\_Final} state and terminates.
Otherwise, policy creation is repeated with the system transiting to \textit{O\_CreatePolicy}.

\begin{minipage}{\linewidth}
\begin{lstlisting}[caption={Preparation}]
   $\land\,\,oState[o]\,\,=\,\,O\_Waiting$
   $\land\,\,IF\,\,Len(policyPool)\,\,>\,\,0\,\,THEN$
     $\land\,\,oBuffer'\,\,=\,\,[\,\,oBuffer\,\,EXCEPT\,\,![o]\,\,=\,\,Head(policyPool)\,\,]$
     $\land\,\,oState'\,\,=\,\,[\,\,oState\,\,EXCEPT\,\,![o]\,\,=\,\,O\_CreatePolicy\,\,]$
     $\land\,\,policyPool'\,\,=\,\,Tail(policyPool)$
     $\land\,\,UNCHANGED\,\,\langle\,\,policySource,\,\,identityBlockchainVars\,\,\rangle$
   $ELSE$
     $\land\,\,oState'\,\,=\,\,[\,\,oState\,\,EXCEPT\,\,![o]\,\,=\,\,O\_Final\,\,]$
     $\land\,\,UNCHANGED\,\,\langle\,\,oBuffer,\,\,tempVars,\,\,identityBlockchainVars\,\,\rangle$
\end{lstlisting}
\end{minipage}

\textit{(ii) CreatePolicy: }As part of this state, the organization defines a consumption or a sharing policy and stores it in the identity blockchain.
The transaction is sent with a \textit{TX\_POLICY} identifier.
Once done, the system transits to the initial state, waiting to create more policies or terminating the process.

\begin{minipage}{\linewidth}
\begin{lstlisting}[caption={Create Policy}]       
   $\land\,\,oState[o]\,\,=\,\,O\_CreatePolicy$
   $\land\,\,blockchainPolicyTxn(o,\,\,oBuffer[o][1],\,\,oBuffer[o][2])$
   $\land\,\,oState'\,\,=\,\,[\,\,oState\,\,EXCEPT\,\,![o]\,\,=\,\,O\_Waiting\,\,]$
   $\land\,\,UNCHANGED\,\,\langle\,\,tempVars,\,\,oBuffer,\,\,iblockTx\,\,\rangle$
\end{lstlisting}
\end{minipage}

\textit{(iii) ConfirmReceipt: }The blockchain system confirms and transfers a valid transaction to blockTx, which is a compilation of verified transactions in the blocks.
In this case, \textit{TX\_POLICY} is a valid transaction that indicates a consumption or sharing policy transaction.

\begin{minipage}{\linewidth}
\begin{lstlisting}[caption={Confirm Receipt}]
   $\exists\,\,receiptTx\,\,\in\,\,itxPool:$
     $\land\,\,receiptTx.type\,\,=\,\,TX\_POLICY$
     $\land\,\,itxPool'\,\,=\,\,itxPool\,\,\backslash\,\,\{receiptTx\}$
     $\land\,\,iblockTx'\,\,=\,\,iblockTx\,\,\cup\,\,\{receiptTx\}$
     $\land\,\,UNCHANGED\,\,\langle\,\,tempVars,\,\,organizationVars\,\,\rangle$
\end{lstlisting}
\end{minipage}

\textit{(iv) Termination: }Once the organization stores all the policies on the identity blockchain, the entire process is terminated. 
Since no further changes are required, termination has no primed variables.

\begin{minipage}{\linewidth}
\begin{lstlisting}[caption={Termination}]
   $\land\,\,\forall\,\,o\,\,\in\,\,Organization\,\,:\,\,oState[o]\,\,=\,\,O\_Final$
   $\land\,\,UNCHANGED\,\,vars$
\end{lstlisting}
\end{minipage}

\begin{figure}[t]
\centering
\includegraphics[width=0.75\textwidth]{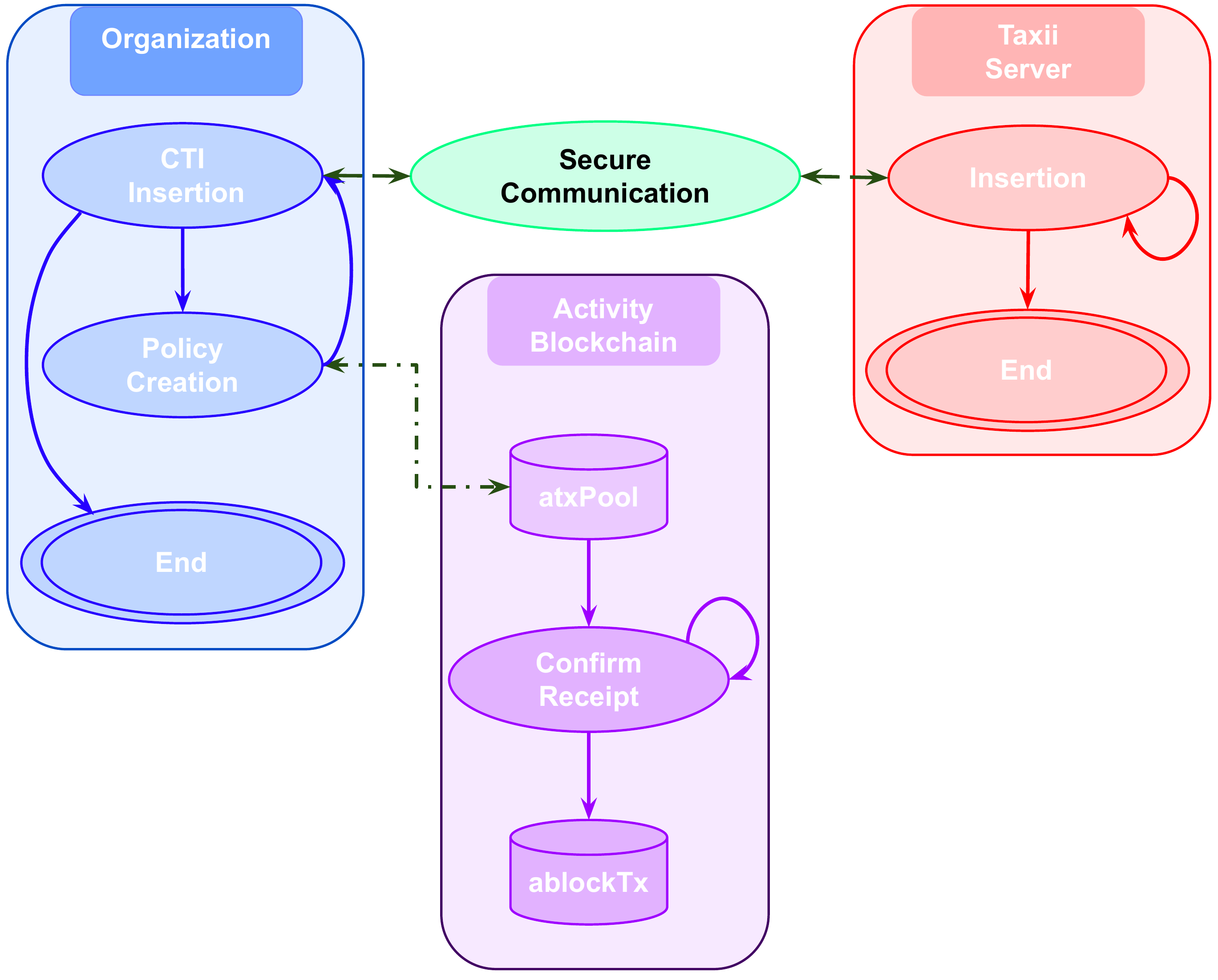}
\caption{CTI information insertion Workflow.}
\label{fig:ctiinsertion}
\end{figure}

\subsection{\label{subsec:ctispec}Cyber threat information insertion specification}Subsequently, we formalize the insertion of cyber threat information into the Taxii server.
The workflow involves interaction between an organization (O), a Taxii server (S), and the activity blockchain (Figure \ref{fig:ctiinsertion}).
InitOrganization, InitServer, and InitActivityBlockchain are the organization, server, and activity blockchain variables' initial values.
The channel variable and state-space specification remain unchanged.
The possible transitions from the initial state are described in the following \textit{State Space}.

The organization, in this case, intends to insert the cyber threat information into the Taxii server.
Further, the CTI is associated with the consumption and sharing policies created on the identity blockchain.
Finally, with the insertion into the blockchain, a notification is sent to the organization's relevant community.
The organization begins with a \textit{O\_Waiting} and then transits through \textit{O\_Insert}, \textit{O\_AssociatePolicy}, and \textit{O\_Final}. In the end, the organization stores the CTI in the Taxii server.

\begin{minipage}{\linewidth}
\begin{lstlisting}[caption={State Space}]
 $Next\,\triangleq$ 
   $\lor\,\,Setup$
   $\lor\,\,\exists\,\,o \in Organization:$
     $\lor\,\,Preparation(o) \lor InsertCyberInformation(o)$
     $\lor\,\,InsertionFailed(o) \lor PolicyAssociation(o)$
   $\lor\,\,\exists\,\,s \in Server$: 
     $\lor\,\,Insertion(s)$
   $\lor\,\,ConfirmReceipt$
   $\lor\,\,Termination$
            
 $Spec\,\triangleq$
   $Init\,\land\,\ast\,[Next]_{vars}\,\land\,WF\_vars(Next)$
\end{lstlisting}
\end{minipage}

\textit{(i) Preparation: }Like the previous one, this state acts as a pre-processing state.
The system checks if there are any more CTI to be inserted into the Taxii server.
If the queue is empty, the system transits to the \textit{O\_Final} state and terminates.
Otherwise, CTI insertion is repeated with the system transiting to \textit{O\_Insert}.

\begin{minipage}{\linewidth}
\begin{lstlisting}[caption={Preparation}]
   $\land\,\,oState[o]\,\,=\,\,O\_Waiting$
   $\land\,\,IF\,\,Len(cyberPool)\,\,>\,\,0\,\,THEN$
     $\land\,\,oBuffer'\,\,=\,\,[\,\,oBuffer\,\,EXCEPT\,\,![o]\,\,=\,\,Head(cyberPool)\,\,]$
     $\land\,\,oState'\,\,=\,\,[\,\,oState\,\,EXCEPT\,\,![o]\,\,=\,\,O\_Insert\,\,]$
     $\land\,\,cyberPool'\,\,=\,\,Tail(cyberPool)$
     $\land\,\,UNCHANGED\,\,\langle\,\,cyberSource,\,\,serverVars,\,\,commVars,\,\,activityBlockchainVars\,\,\rangle$
   $ELSE$
     $\land\,\,oState'\,\,=\,\,[\,\,oState\,\,EXCEPT\,\,![o]\,\,=\,\,O\_Final\,\,]$
     $\land\,\,UNCHANGED\,\,\langle\,\,tempVars,\,\,oBuffer,\,\,serverVars,\,\,commVars,\,\,activityBlockchainVars\,\,\rangle$
\end{lstlisting}
\end{minipage}

\textit{(ii) InsertCyberInformation: }An organization initiates the CTI creation process by sending the message containing the details of a threat to the Taxii server.
Once the message is sent, the system transits to \textit{O\_AssociatePolicy} state.
The organization then waits for the response from the Taxii server.

\begin{minipage}{\linewidth}
\begin{lstlisting}[caption={Insert Cyber Information}]         
 $\land\,\,oState[o]\,\,=\,\,O\_Insert$
 $\land\,\,\exists\,\,s \in Server:$
   $\land\,\,sState[s]\,\,=\,\,S\_Waiting$
   $\land\,\,LET$ 
     $m\,\,==\,\,oBuffer[o]$
   $IN$ 
     $Send([src\,\,\mapsto\,\,o,\,\,dst\,\,\mapsto\,\,s,\,\,type\,\,\mapsto\,\,M\_Insertion,\,\,data\,\,\mapsto\,\,\langle\,\,o,\,\,m[1],\,\,m[2]\,\,\rangle])$
 $\land\,\,oState'\,\,=\,\,[\,\,oState\,\,EXCEPT\,\,![o]\,\,=\,\,O\_AssociatePolicy\,\,]$
 $\land\,\,UNCHANGED\,\,\langle\,\,tempVars,\,\,oBuffer,\,\,serverVars,\,\,activityBlockchainVars\,\,\rangle$
\end{lstlisting}
\end{minipage}

\textit{(iii) InsertionFailed: }If the Taxii server fails to insert the CTI sent by the organization, the server responds with a failure message back to the organization. 
The server remains in \textit{S\_Waiting} state to process further requests.
On receiving the failure message, the system transits to \textit{O\_Final} state and terminates.

\begin{minipage}{\linewidth}
\begin{lstlisting}[caption={Insertion Failed}]
   $\exists\,\,c \in channel:$
     $\land\,\,c.dst\,\,=\,\,o$
     $\land\,\,c.type\,\,=\,\,M\_NO$
     $\land\,\,oState'\,\,=\,\,[\,\,oState\,\,EXCEPT ![o]\,\,=\,\,O\_Final\,\,]$
     $\land\,\,UNCHANGED\,\,\langle\,\,tempVars,\,\,oBuffer,\,\,serverVars,\,\,commVars,\,\,activityBlockchainVars\,\,\rangle$ 
\end{lstlisting}
\end{minipage}

\textit{(iv) PolicyAssociation: }On receiving a message indicating successful insertion of CTI in the Taxii server, the organization sends a transaction to the activity blockchain, associating the consumption and sharing policy with the newly inserted CTI.
The association transaction is identified by \textit{ATX\_CYBER\_THREAT}.

\begin{minipage}{\linewidth}
\begin{lstlisting}[caption={Policy Association}]
   $\exists\,\,c \in channel:$
     $\land\,\,c.dst\,\,=\,\,o$
     $\land\,\,c.type\,\,=\,\,M\_OK$
     $\land\,\,oState'\,\,=\,\,[\,\,oState\,\,EXCEPT ![o]\,\,=\,\,O\_Waiting\,\,]$
     $\land\,\,LET$ 
       $m\,\,==\,\,oBuffer[o]$
       $pid\,\,==\,\,POLICY\_ID$
     $IN$ 
       $blockchainPolicyTxn(o,\,\,m[1],\,\,pid)$
       $UNCHANGED\,\,\langle\,\,tempVars,\,\,oBuffer,\,\,serverVars,\,\,commVars,\,\,ablockTx\,\,\rangle$
\end{lstlisting}
\end{minipage}

The Taxii server is responsible for storing the CTI information. 
The server remains in \textit{S\_Waiting} state, waiting for insertion request from an organization.

\begin{minipage}{\linewidth}
\begin{lstlisting}[caption={Insertion}]
   $\exists\,\,c \in channel:$
     $\land\,\,c.dst\,\,=\,\,s$
     $\land\,\,c.type\,\,=\,\,M\_Insertion$
     $\land\,\,sState[s]\,\,=\,\,S\_Waiting$
     $\land\,\,IF\,\,InsertCTI(c.data[2],\,\,c.data[3])\,\,THEN$
       $\land\,\,RecvInadditionSend(c,\,\,[src\,\,\mapsto\,\,s,\,\,dst\,\,\mapsto\,\,c.data[1],\,\,type\,\,\mapsto\,\,M\_OK,\,\,data\,\,\mapsto\,\,\langle\,\,\rangle])$
       $\land\,\,UNCHANGED\,\,\langle\,\,tempVars,\,\,organizationVars,\,\,serverVars,\,\,activityBlockchainVars\,\,\rangle$
     $ELSE$
       $\land\,\,RecvInadditionSend(c,\,\,[src\,\,\mapsto\,\,s,\,\,dst\,\,\mapsto\,\,c.data[1],\,\,type\,\,\mapsto\,\,M\_NO,\,\,data\,\,\mapsto\,\,\langle\,\,\rangle])$
       $\land\,\,UNCHANGED\,\,\langle\,\,tempVars,\,\,organizationVars,\,\,serverVars,\,\,activityBlockchainVars\,\,\rangle$
\end{lstlisting}
\end{minipage}

\textit{(v) Insertion: }In this state, the Taxii server inserts the details received from an organization. 
If the insertion is successful, the server returns a success message.
Otherwise, the server returns a failed message.
During the process, the server remains in \textit{S\_Waiting} state.

\textit{(vi) ConfirmReceipt: }The activity blockchain, similar to identity blockchain, confirms and transfers a valid transaction to blockTx, which is a compilation of verified transactions in the blocks.
In this case, \textit{ATX\_CYBER\_THREAT} is a valid transaction that indicates a CTI record with an associated consumption and sharing policy transaction.

\begin{minipage}{\linewidth}
\begin{lstlisting}[caption={Confirm Receipt}]
   $\exists\,\,receiptTx\,\,\in\,\,atxPool:$
     $\land\,\,receiptTx.type\,\,=\,\,ATX\_CYBER\_THREAT$
     $\land\,\,atxPool'\,\,=\,\,atxPool\,\,\backslash\,\,\{receiptTx\}$
     $\land\,\,ablockTx'\,\,=\,\,ablockTx\,\,\cup\,\,\{receiptTx\}$
     $\land\,\,UNCHANGED\,\,\langle\,\,tempVars,\,\,organizationVars,\,\,serverVars,\,\,commVars\,\,\rangle$
\end{lstlisting}
\end{minipage}

\textit{(vii) Termination: }Once the organization associates the policies with the CTI on the activity blockchain, the entire process is terminated. 
Since no further changes are required, termination has no primed variables.

\begin{minipage}{\linewidth}
\begin{lstlisting}[caption={Termination}]
   $\land\,\,\forall\,\,o\,\,\in\,\,Organization\,\,:\,\,oState[o]\,\,=\,\,O\_Final$
   $\land\,\,UNCHANGED\,\,vars$
\end{lstlisting}
\end{minipage}

\begin{figure}[t]
\centering
\includegraphics[width=0.8\textwidth]{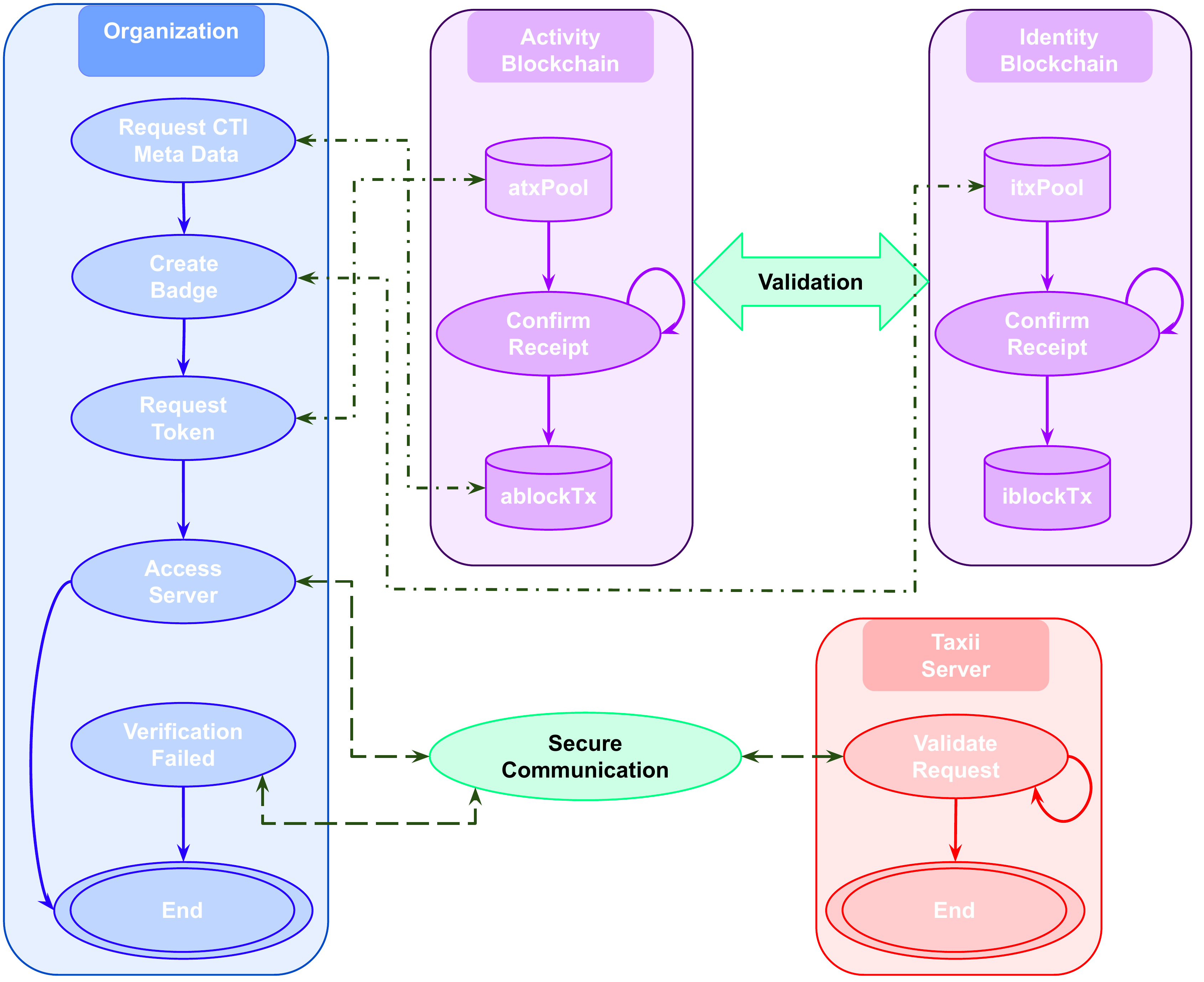}
\caption{Access authorization workflow.}
\label{fig:accessauthorization}
\end{figure}

\subsection{\label{subsec:accessspec}Access authorization specification}Finally, we formalize the access authorization process initiated by a consumer.
The workflow involves interaction between a consumer (C), a Taxii server (S), the activity blockchain, and the identity blockchain (Figure \ref{fig:accessauthorization}).
InitConsumer, InitServer, InitActivityBlockchain, and InitIdentityBlockchain are the consumer, server, activity blockchain, and identity blockchain variables' initial values.
The channel variable and state-space specification remain unchanged.
The possible transitions from the initial state are described in the following equation:

\begin{minipage}{\linewidth}
\begin{lstlisting}[caption={State Space}]
 $Next\,\triangleq$ 
   $\lor\,\,Setup$
   $\lor\,\,\exists\,\,c \in Consumer:$
     $\lor\,\,Preparation(c) \lor RequestMetaData(c)$
     $\lor\,\,CreateBadge(c) \lor RequestToken(c)$
     $\lor\,\,AccessServer(c) \lor VerificationFailed(c)$
     $\lor\,\,AccessCTI(c)$
   $\lor\,\,\exists\,\,s \in Server$: 
     $\lor\,\,ValidateRequest(s)$
   $\lor\,\,IConfirmReceipt$
   $\lor\,\,AConfirmReceipt$
   $\lor\,\,Termination$
            
 $Spec\,\triangleq$
   $Init\,\land\,\ast\,[Next]_{vars}\,\land\,WF\_vars(Next)$
\end{lstlisting}
\end{minipage}

The consumer is interested in accessing CTI present in the Taxii server. 
The consumer begins with a \textit{C\_Waiting} and then transits through \textit{C\_RequestMetaData}, \textit{C\_CreateBadge}, \textit{C\_RequestToken}, \textit{C\_AccessServer}, \textit{C\_ValidateAccess}, \textit{C\_Final}. 
In the end, the consumer can either access the CTI record or the access request is rejected.

\textit{(i) RequestMetaData: }The consumer begins searching and accessing the metadata of the CTI on the activity blockchain.
The consumer then saves the information to be used for further requests.
Once done, the system transits to \textit{C\_CreateBadge} where the consumer creates a badge for requesting the access token.

\begin{minipage}{\linewidth}
\begin{lstlisting}[caption={Request Meta Data}]
   $\exists\,\,tx \in ablockTx:$
     $\land\,\,cState[c]\,\,=\,\,C\_RequestMetaData$
     $\land\,\,tx.type\,\,=\,\,ATX\_CYBER\_THREAT$
     $\land\,\,tx.who\,\,=\,\,ORGANIZATION$
     $\land\,\,ctiInfo'\,\,=\,\,[\,\,ctiInfo\,\,EXCEPT ![c]\,\,=\,\,tx.data\,\,]$
     $\land\,\,cState'\,\,=\,\,[\,\,cState\,\,EXCEPT\,\,![c]\,\,=\,\,C\_CreateBadge\,\,]$
     $\land\,\,UNCHANGED\,\,\langle\,\,cBuffer,\,\,accessPool,\,\,accessSource,\,\,serverVars,$
     $commVars,\,\,identityBlockchainVars,\,\,activityBlockchainVars\,\,\rangle$
\end{lstlisting}
\end{minipage}

\textit{(ii) CreateBadge: }Next, the consumer creates the badge on the identity blockchain.
The \textit{blockchainIdentityTxn} function abstracts the process of sending a transaction to the identity blockchain.
Once the badge is created, the system transits to \textit{C\_RequestToken} state.

\begin{minipage}{\linewidth}
\begin{lstlisting}[caption={Create Badge}]
   $\land\,\,cState[c]\,\,=\,\,C\_CreateBadge$
   $\land\,\,LET$
     $bgid\,\,==\,\,BADGEID$
     $badge\,\,==\,\,BADGE$
   $IN$
     $blockchainIdentityTxn(c,\,\,bgid,\,\,badge)$
   $\land\,\,cState'\,\,=\,\,[\,\,cState\,\,EXCEPT\,\,![c]\,\,=\,\,C\_RequestToken\,\,]$
   $\land\,\,UNCHANGED\,\,\langle\,\,ctiInfo,\,\,iblockTx,\,\,cBuffer,\,\,accessPool,\,\,accessSource,$
   $\,\,serverVars,\,\,commVars,\,\,activityBlockchainVars\,\,\rangle$
\end{lstlisting}
\end{minipage}

\textit{(iii) RequestToken: }Once the badge is created, the consumer requests the access token required to access the CTI by presenting the badge created in the previous setup.
The access token request is sent to the activity blockchain, which validates the badge details by requesting verification from the identity blockchain.
On successful validation, the access token is created on the activity blockchain, and its reference is returned to the consumer.
The system then transits to \textit{C\_AccessServer} state.
In case the validation fails, the system transits to \textit{C\_Final} and the system terminates.

\begin{minipage}{\linewidth}
\begin{lstlisting}[caption={Request Token}]
   $\land\,\,cState[c]\,\,=\,\,C\_RequestToken$
   $\land\,\,LET$
     $bgid\,\,==\,\,BADGEID$
   $IN$
     $IF\,\,ValidateBadge(bgid)\,\,THEN$
       $\land\,\,blockchainActivityTxn(c,\,\,bgid)$
       $\land\,\,cState'\,\,=\,\,[\,\,cState\,\,EXCEPT\,\,![c]\,\,=\,\,C\_AccessServer\,\,]$
       $\land\,\,UNCHANGED\,\,\langle\,\,ctiInfo,\,\,ablockTx,\,\,cBuffer,\,\,accessPool,\,\,accessSource,$
       $\,\,serverVars,\,\,commVars,\,\,identityBlockchainVars\,\,\rangle$
     $ELSE$
       $\land\,\,cState'\,\,=\,\,[\,\,cState\,\,EXCEPT\,\,![c]\,\,=\,\,C\_Final\,\,]$
       $\land\,\,UNCHANGED\,\,\langle\,\,ctiInfo,\,\,cBuffer,\,\,accessPool,\,\,accessSource,$
       $\,\,serverVars,\,\,commVars,\,\,activityBlockchainVars,\,\,identityBlockchainVars\,\,\rangle$
\end{lstlisting}
\end{minipage}

\textit{(iv) AccessServer: }With the token reference issued on the activity blockchain, the consumer requests access to the CTI identified by the id \textit{CID}.
The system then transits to \textit{C\_ValidateAccess} state where the Taxii server validates the access token.

\begin{minipage}{\linewidth}
\begin{lstlisting}[caption={Access Server}]    
   $\exists\,\,s \in Server:$
     $\land\,\,cState[c]\,\,=\,\,C\_AccessServer$
     $\land\,\,LET$ 
       $token\,\,==\,\,TOKEN$
       $cid\,\,==\,\,CID$
     $IN$
       $Send([src\,\,\mapsto\,\,c,\,\,dst\,\,\mapsto\,\,s,\,\,type\,\,\mapsto\,\,M\_Access,\,\,data\,\,\mapsto\,\,\langle\,\,token,\,\,cid\,\,\rangle])$
   $\land\,\,cState'\,\,=\,\,[\,\,cState\,\,EXCEPT\,\,![c]\,\,=\,\,C\_ValidateAccess\,\,]$
   $\land\,\,UNCHANGED\,\,\langle\,\,ctiInfo,\,\,ablockTx,\,\,cBuffer,\,\,accessPool,\,\,accessSource,$
   $\,\,serverVars,\,\,atxPool,\,\,identityBlockchainVars\,\,\rangle$
\end{lstlisting}
\end{minipage}

\textit{(v) VerificationFailed: }The system transits to this state if the token verification fails.
The system then transits to \textit{C\_Final} state, and the system terminates.

\begin{minipage}{\linewidth}
\begin{lstlisting}[caption={Verification Failed}]
   $\exists\,\,ch \in channel:$
     $\land\,\,ch.dst\,\,=\,\,c$
     $\land\,\,ch.type\,\,=\,\,M\_NO$
     $\land\,\,cState'\,\,=\,\,[\,\,cState\,\,EXCEPT\,\,![c]\,\,=\,\,C\_Final\,\,]$
     $\land\,\,UNCHANGED\,\,\langle\,\,tempVars,\,\,cBuffer,\,\,serverVars,\,\,commVars,$
     $\,\,identityBlockchainVars,\,\,activityBlockchainVars\,\,\rangle$ 
\end{lstlisting}
\end{minipage}

\textit{(vi) AccessCTI: }The system transits to this state if the token verification succeeds.
The system then allows the consumer to access the CTI record, and the system transits to \textit{C\_Final} state and terminates.

\begin{minipage}{\linewidth}
\begin{lstlisting}[caption={Access CTI}]
   $\exists\,\,ch \in channel:$
     $\land\,\,ch.dst\,\,=\,\,c$
     $\land\,\,ch.type\,\,=\,\,M\_OK$
     $\land\,\,cState'\,\,=\,\,[\,\,cState\,\,EXCEPT\,\,![c]\,\,=\,\,C\_Final\,\,]$
     $\land\,\,UNCHANGED\,\,\langle\,\,tempVars,\,\,cBuffer,\,\,serverVars,\,\,commVars,$
     $\,\,identityBlockchainVars,\,\,activityBlockchainVars\,\,\rangle$     
\end{lstlisting}
\end{minipage}

The Taxii server is responsible for validating the access token sent by the consumer with the access request. 
The server remains in \textit{S\_Waiting} state, waiting for other access requests from other consumers.

\textit{(vii) ValidateRequest: }In this state, the Taxii server validates the consumer's access request based on the access token reference sent by the consumer. 
The verification process is abstracted by the \textit{VerifyAccess} function.
If the verification is successful, the server returns a success message.
Otherwise, the server returns a failed message.
During the process, the server remains in \textit{S\_Waiting} state.

\begin{minipage}{\linewidth}
\begin{lstlisting}[caption={Validate Request}]
   $\exists\,\,c \in channel:$
     $\land\,\,c.dst\,\,=\,\,s$
     $\land\,\,c.type\,\,=\,\,M\_Access$
     $\land\,\,sState[s]\,\,=\,\,S\_Waiting$
     $\land\,\,IF\,\,VerifyAccess(c.data[1],\,\,c.data[2])\,\,THEN$
       $RecvInadditionSend(c,\,\,[src\,\,\mapsto\,\,s,\,\,dst\,\,\mapsto\,\,c.src,\,\,type\,\,\mapsto\,\,M\_OK,\,\,data\,\,\mapsto\,\,\langle\,\,\rangle])$
     $ELSE$
       $RecvInadditionSend(c,\,\,[src\,\,\mapsto\,\,s,\,\,dst\,\,\mapsto\,\,c.src,\,\,type\,\,\mapsto\,\,M\_NO,\,\,data\,\,\mapsto\,\,\langle\,\,\rangle])$
     $\land\,\,sState'\,\,=\,\,[\,\,sState\,\,EXCEPT\,\,![s]\,\,=\,\,S\_Waiting\,\,]$
     $\land\,\,UNCHANGED\,\,\langle\,\,tempVars,\,\,consumerVars,\,\,sBuffer,$
     $\,\,identityBlockchainVars,\,\,activityBlockchainVars\,\,\rangle$
\end{lstlisting}
\end{minipage}

\textit{(viii) IConfirmReceipt: }\textit{ITX\_BADGE} is a valid transactions that indicates the badge creation on the identity blockchain.
The consumer then uses the created badges to request the access token.

\begin{minipage}{\linewidth}
\begin{lstlisting}[caption={Confirm Receipt for Identity Blockchain}]
   $\exists\,\,receiptTx\,\,\in\,\,itxPool:$
     $\land\,\,receiptTx.type\,\,=\,\,ITX\_BADGE$
     $\land\,\,itxPool'\,\,=\,\,itxPool\,\,\backslash\,\,\{receiptTx\}$
     $\land\,\,iblockTx'\,\,=\,\,iblockTx\,\,\cup\,\,\{receiptTx\}$
     $\land\,\,UNCHANGED\,\,\langle\,\,consumerVars,\,\,serverVars,\,\,commVars,\,\,accessPool,$
     $accessSource,\,\,ctiInfo,\,\,activityBlockchainVars\,\,\rangle$
\end{lstlisting}
\end{minipage}

\textit{(ix) AConfirmReceipt: }In this case, \textit{ATX\_ACCESS\_TOKEN} is a valid transaction that indicates a CTI access token granted to the consumer to access the CTI record on the Taxii server.

\begin{minipage}{\linewidth}
\begin{lstlisting}[caption={Confirm Receipt for Activity Blockchain}]
   $\exists\,\,receiptTx\,\,\in\,\,atxPool:$
     $\land\,\,receiptTx.type\,\,=\,\,ATX\_ACCESS\_TOKEN$
     $\land\,\,atxPool'\,\,=\,\,atxPool\,\,\backslash\,\,\{receiptTx\}$
     $\land\,\,ablockTx'\,\,=\,\,ablockTx\,\,\cup\,\,\{receiptTx\}$
     $\land\,\,UNCHANGED\,\,\langle\,\,consumerVars,\,\,serverVars,\,\,commVars,\,\,accessPool,$
     $accessSource,\,\,ctiInfo,\,\,identityBlockchainVars\,\,\rangle$
\end{lstlisting}
\end{minipage}

\textit{(x) Termination: }Once the consumer request to access a CTI is either granted or rejected, and the entire process is terminated. 
Since no further changes are required, termination has no primed variables.

\begin{minipage}{\linewidth}
\begin{lstlisting}[caption={Termination}]
   $\land\,\,\forall\,\,o\,\,\in\,\,Organization\,\,:\,\,oState[o]\,\,=\,\,O\_Final$
   $\land\,\,UNCHANGED\,\,vars$
\end{lstlisting}
\end{minipage}

\subsection{\label{subsec:modelchecking}Model checking}
The TLC model evaluation involves parsing the state space to find critical states, e.g., states leading to deadlock conditions or violation of invariants.
We performed simulation on the TLA+ model verification tool to check the proposed system's correctness and the critical conditions.
We ran the experiment on a Windows 10-based AMD Ryzen-7 octa-core machine with 16 GB RAM.
The evaluation is performed for each of the workflows.
The results are presented in Table~\ref{tab:simulation}.

The table lists the different workflows involved in the proposal, from the organization registering to share the CTI to the access authorization workflow.
In case there are failures in the workflow due to validation error, a separate entry is added to the table with relevant details.
The table lists the details like depth in the state-space the simulation tool went to, the number of states discovered, the number of distinct states out of those, and the errors encountered. 
We found that the system always resulted in a consistent and safe state.
Based on the results presented in the Table~\ref{tab:simulation}, we can conclude that the proposed scheme is error-free.

\begin{table}[b]
\centering
\footnotesize
\caption{TLA+ simulation results.}
\label{tab:simulation}
\begin{tabular}{c|c|c|c|c} 
\hline
\textbf{Situation} & \textbf{Depth} & \textbf{States Found} & \textbf{Distinct} & \textbf{Errors} \\  
& & & \textbf{States} & \\  

\hline
\hline
Registration & 13 & 57 & 25 & 0 \\
\hline
Registration (failure) & 6 & 8 & 6 & 0 \\
\hline
Policy creation & 14 & 660 & 287 & 0 \\
\hline
CTI creation & 21 & 8,547 & 1,922 & 0 \\
\hline
Access authorization & 11 & 44 & 22 & 0 \\
\hline
Access authorization & 7 & 11 & 8 & 0 \\
(badge verification failure) & & & & \\
\hline
\end{tabular}
\end{table}
\section{\label{sec:discussion}Discussion}
In this section, we evaluate the privacy and accountability guarantees of the proposed solution.
Next, we present the performance evaluation of the proposed solution and identify its overheads. 
We suggest quantitative measures to reduce the overhead.
Finally, we describe additional blockchain mechanisms to improve the sharing of CTI.

\subsection{Security analysis}
\label{sec:secanalysis}
Blockchain-based access control provides organizations with complete control over their data.
The blockchain ensures that only organizations that comply with the sharing policy can access the data.
We evaluate the privacy and accountability guarantees of the proposed framework using the following hypotheses.

\noindent \textbf{Hypothesis H1:} It is impossible to correlate between profile and activity without compromising the whole network.
Based on the threat sharing framework, the profile by itself does not expose any sensitive information. 
The mapping between the profile and the {\CTINet} activity is required to extract legible information about the sharing organization.
Thus, a malicious actor needs to either expose the correlation between profile and {\CTINet} activity.
To achieve this, he/she needs to compromise both the {\AdminNet} and the {\CTINet} as with access to the {\AdminNet} one can access the profile but not the activity while with access to the {\CTINet} one can only see the activity but no profile information.

Access to {\AdminNet} is regulated by the consortium of registrars who store the mapping between the profile and the pseudo-identity on the {\AdminNet} (refer Table \ref{tab:allowtran}).
Thus, to compromise the {\AdminNet}, one needs to compromise a two-thirds majority of registrars.
This is not easy to achieve as the registrars have a stake in the platform's proper functioning.
Also, access to {\CTINet} is restricted by the blockchain-based access control.

\noindent \textbf{Hypothesis H2:} A misbehaving organization can be identified and exposed. 
The {\CTINet} logs the activity of an organization on the {\CTINet} using the {\tuid}.
Thus, the sharing organization can audit the activity on their data. 
In case of unauthorized access, the sharing organization can report the {\tuid} of the consuming organization.
The mapping between the real identity and {\tuid} exists on the {\AdminNet}.
Thus, the consortium of registrars can reach a consensus to expose the misbehaving organization's real identity.

\subsection{Performance}
\label{sec:perf}
The main overhead of decoupling the {\AdminNet} and the {\CTINet} is profile badge (PB) generation on the {\AdminNet} (see stage 1 in the authorization process~\ref{subsec:authorization}).
The activity on the {\CTINet} is similar to other blockchain-based access control solutions (except the queries for the PB from the {\AdminNet}).
Assuming that a user uses $N$ different temporary {\tuid} and on an average user use $k$ PBs between profile changes, the user generates $k*N$ new badges.
To minimize the badge generation overhead, we can minimize the value of $k$ and $N$.
To reduce the number of PBs, we should aim to reuse them PBs.
This is why we suggest that organizations should reuse other organizations' policies.
Also, to reduce the value of $N$, we need to establish a trade-off between privacy (num of temporary {\tuid}) and overhead ($k*N$).

\subsection{Additional blockchain mechanisms}
\label{sec:blockmechanism}
\noindent \textbf{Reputation contract (RPC).}
This contract maps each organization's blockchain address identity with a reputation profile. 
The reputation profile can be a single reputation score or can be a complex multi-dimensional profile. 
The RC determines an organization's reputation profile according to the reputation rates of other organizations in the network.  
There are many reputation systems in the literature (e.g., \cite{hendrikx2015reputation, ghasempouri2019modeling}) applicable for the proposed solution.  
By integrating this contract with the system, an organization can incorporate organizations' reputation profile within their consumption and sharing policy. 
For example, organizations can define rules, like only organizations with a reputation score higher than some threshold are allowed to see their data. 
Alternatively, an organization can share information with top analysts. 

\noindent \textbf{Automation via legal contract.}
In many cases, threat sharing may involve a legal agreement between the CTI producer and consumer. 
Today, such an agreement requires a pre-established relationship between the CTI producer and consumer, which contradicts the {\methodName} concept of exchanging threat information anonymously and on the fly. 
To address this challenge, a legal contract mechanism can be incorporated in {\methodName}. 
This blockchain mechanism regulates the fulfillment of legal agreements regarding CTI exchange. 
When sharing information, the CTI producer can tag CTI with a legal contract specifying the term of use the data consumer must agree before consuming the subject CTI. 
This legal contract ensures that a digital signature of any consumer is stored on the blockchain ledger. 
We suggest leveraging the information exchange policy framework~\cite{iep} for formalizing the term of use. 

\noindent \textbf{Encouraging threat sharing (fairness).}
When threat intelligence is shared anonymously, and organizations can opt into information-sharing coalitions at will, how might a sharing system avoid the potential "free loading" problem? 
What incentive is there for a threat analyst to reciprocate and share data in a world where no one knows whether or not they are sharing? {\methodName} solves this problem by rewarding karma for information shared. 
To receive and consume threat intelligence data, organizations must spend the karma they have earned. 
Moreover, if organizations are incentivized to contribute threat intelligence to consume threat intelligence, what prevents them from anonymously contributing data of questionable quality? 
In {\methodName}, each network member might gain or lose a reputation based on the quality of the data they contribute. 
When a member spends karma to receive threat intelligence data that turns out to be flawed or misrepresented, they can rate the contribution.
Organizations can earn karma discounts usable for future threat intelligence acquisition by providing a rating upon the information they have received. 
If an organization's reputation falls too low because it has repeatedly submitted low-quality data, it will not access critical threat intelligence. 
This feature of TRADE ensures data reliability by isolating non-contributors.
In this, {\methodName} creates a non-monetary, fiduciary data exchange resembling a market based on rewarded goodwill.
\section{Conclusions and Future Work}
\label{sec:conclusions}
In this paper, we presented TRADE, a blockchain-based platform for sharing cyber-threat intelligence across a community of diverse organizations. 
The proposed solution provides a peer-to-peer, privacy-preserving, transparent, and accountable threat sharing network where the information producer has complete control over her threat information. 
TRADE provides a distributed network that is not relying on a single trusted third party. 
TRADE can be easily integrated within existing threat sharing standards (such as TAXII and OpenDXL), thereby facilitating easy technology adaptation within the current organizations' threat sharing workflow.  
This paper aims to trigger collaboration among academic and industrial partners to make TRADE a foundation for a community-based next-generation platform for threat sharing.

\Urlmuskip=0mu plus 1mu\relax
\bibliographystyle{unsrt}  
\bibliography{sections/biblio}

\newpage
\appendix
\section*{\label{sec:appendix}Appendix}
\begin{figure}[h!]
\centering
\includegraphics[scale=0.2]{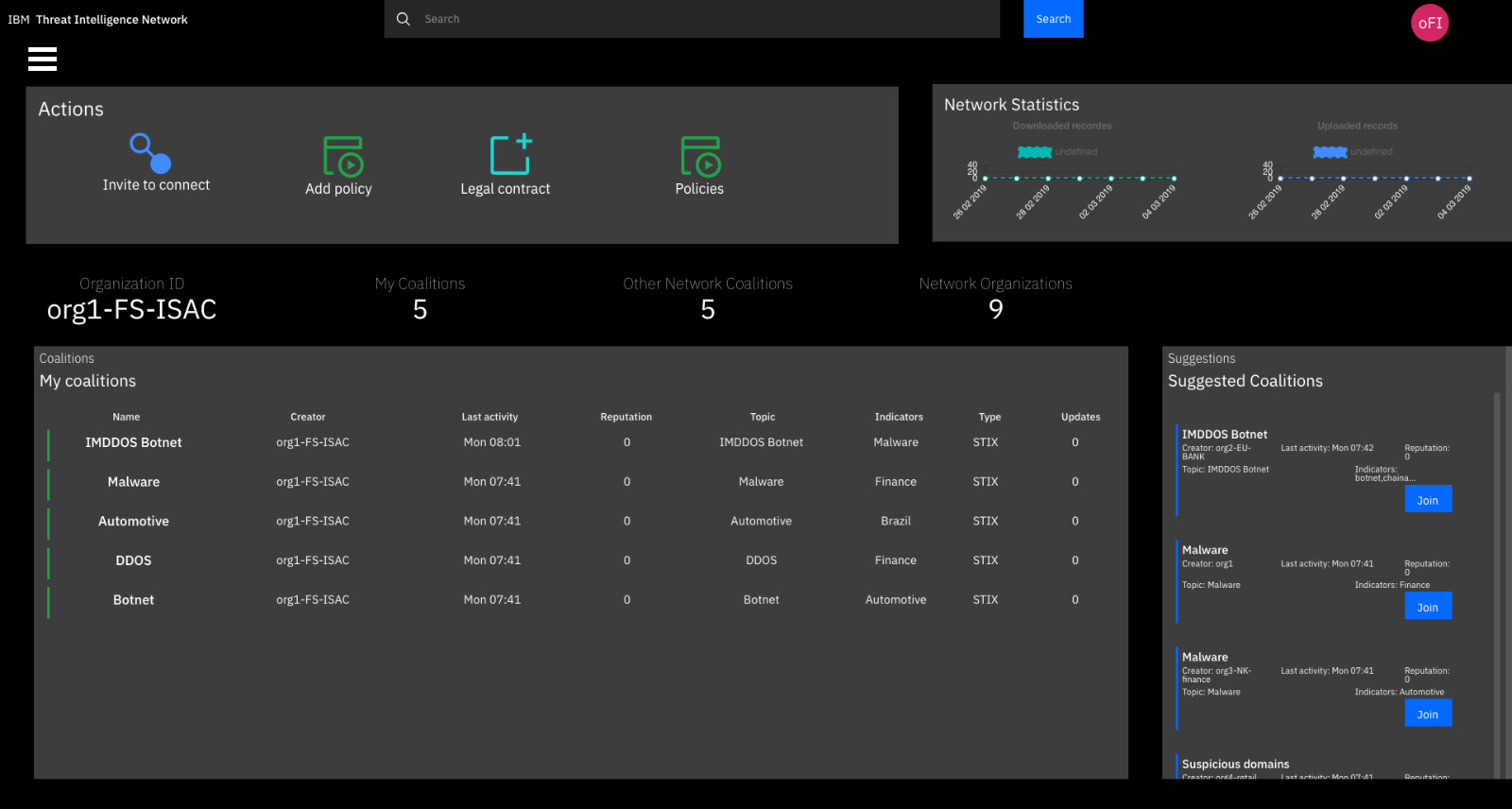}
\caption{TRADE user interface.}
\label{fig:userinterface}
\end{figure}

\begin{figure}[h!]
\centering
\includegraphics[scale=0.2]{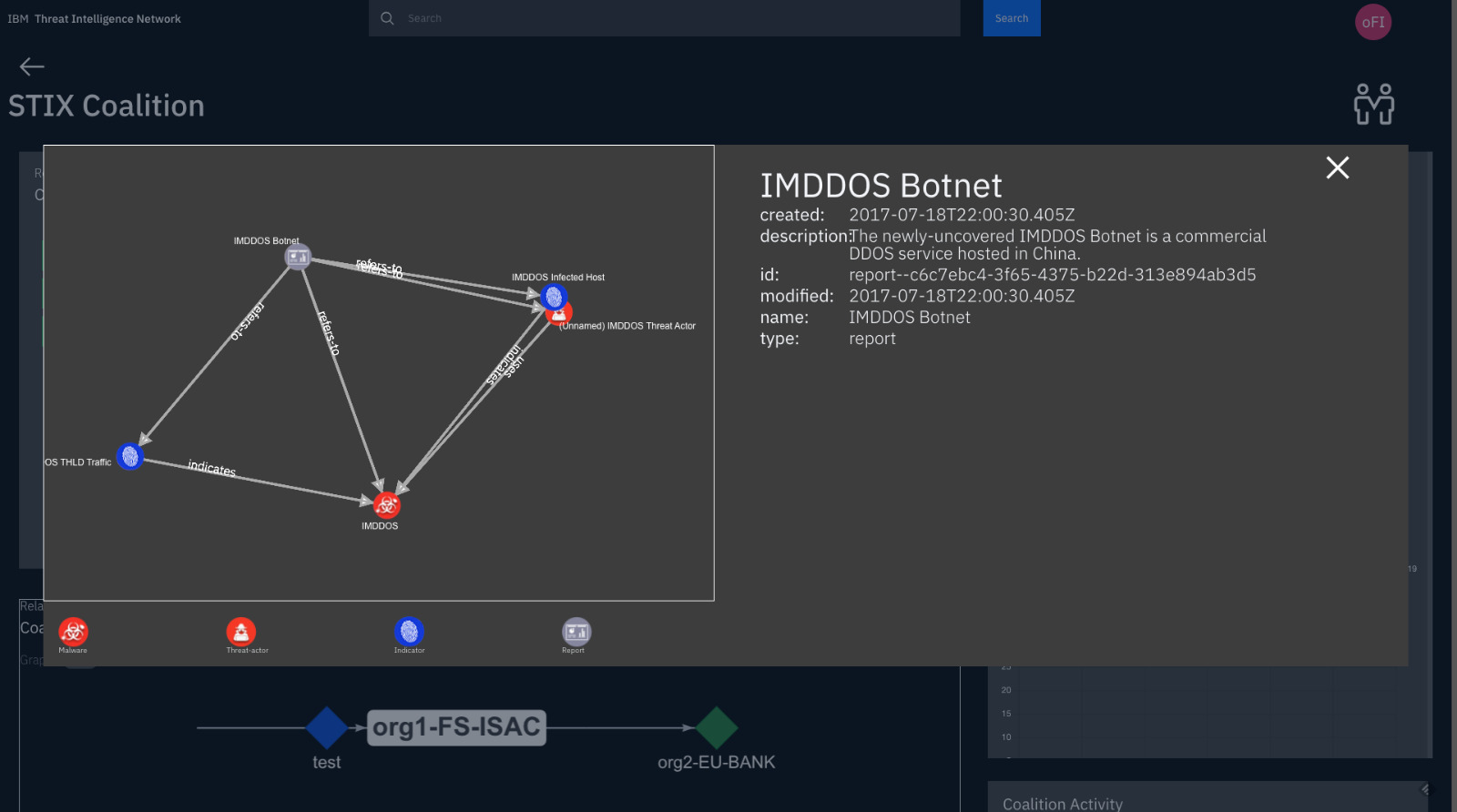}
\caption{Collaborative STIX report on TRADE user interface.}
\label{fig:stix}
\end{figure}

Two representative snapshots from a running instance of the TRADE implementation are shown herein.  
Figure \ref{fig:userinterface} captures the TRADE dashboard and Figure \ref{fig:stix} captures a collaborative STIX report shown on TRADE's user interface.

\end{document}